\documentclass[twocolumn,floatfix,nofootinbib, superscriptaddress, tightenlines,nobibnotes,aps,prd]{revtex4-1}
\usepackage{hyperref}
\usepackage{graphicx}
\usepackage{amsmath,amssymb,latexsym}
\usepackage{delarray}
\usepackage{dcolumn}
\usepackage{bm}
\usepackage{color}
\usepackage{subfigure}
\usepackage{xspace}
\usepackage{breakurl}

\usepackage{mathabx}
\usepackage{mathrsfs}
\usepackage[usenames,dvipsnames]{xcolor}

\definecolor{rossoCP3}{cmyk}{0,.88,.77,.40}
\definecolor{darkBlue}{rgb}{0, 0, 0.8}

\hypersetup{
    bookmarks=true,         
    bookmarksopen=true,     
    bookmarksopenlevel=1,
    unicode=false,          
    pdftoolbar=true,        
    pdfmenubar=true,        
    colorlinks=true,        
    linkcolor=darkBlue,     
    citecolor=darkBlue,        
    filecolor=darkBlue,      
    urlcolor=darkBlue           
}

\def\EeV{\ifmmode {\mathrm{Ee\kern -0.07em V}}\else
                   \textrm{Ee\kern -0.07em V}\fi\xspace}
\def\GeV{\ifmmode {\mathrm{Ge\kern -0.07em V}}\else
                   \textrm{Ge\kern -0.07em V}\fi\xspace}
\def\TeV{\ifmmode {\mathrm{Te\kern -0.07em V}}\else
                   \textrm{Te\kern -0.07em V}\fi\xspace}
\def\PeV{\ifmmode {\mathrm{Pe\kern -0.07em V}}\else
                   \textrm{Pe\kern -0.07em V}\fi\xspace}
\def\eV{\ifmmode {\mathrm{e\kern -0.07em V}}\else
                   \textrm{e\kern -0.07em V}\fi\xspace}
\def\meV{\ifmmode {\mathrm{me\kern -0.07em V}}\else
                   \textrm{me\kern -0.07em V}\fi\xspace}
\def\gcm{\ensuremath{\mathrm{g/cm}^2}\xspace}
\def\Xmax{\ensuremath{X_\mathrm{max}}\xspace}

\def\Sibyll{\textsc{Sibyll2.1}\xspace}
\def\Epos{\textsc{Epos-LHC}\xspace}
\def\QgII{\textsc{QGSJetII-04}\xspace}

\newcommand{\energy}[1]{\ensuremath{10^{#1}}\,\eV}

\newcommand{\CRP} {{\scshape CRPropa}\xspace}
\newcommand{\calQ} {{\cal Q}\xspace}
\newcommand{\Astar} {A^{*}\xspace}
\newcommand{\bolddot}[1] { \overset{\bm .}{#1}}

\makeatletter

\begin{document}

\title{\texorpdfstring{\color{rossoCP3}}{}Origin of the ankle in the ultrahigh energy
  cosmic ray spectrum,\\ and of the extragalactic protons below it}

\author{Michael Unger}
\email{mu495@nyu.edu}
\affiliation{Center for Cosmology and Particle Physics, Department of Physics, New York University, NY 10003, USA
}
\affiliation{Karlsruher Institut f\"ur Technologie, Institut f\"ur
  Kernphysik, Postfach 3640, 76021 Karlsruhe, Germany
}

\author{Glennys R. Farrar}
\email{gf25@nyu.edu}
\affiliation{Center for Cosmology and Particle Physics, Department of Physics, New York University, NY 10003, USA
}

\author{Luis A. Anchordoqui}
\email{laa410@nyu.edu}
\affiliation{Department of Physics and Astronomy, Lehman College, City University of
  New York, NY 10468, USA
}
\affiliation{Department of Physics, Graduate Center, City University
  of New York, 365 Fifth Avenue, NY 10016, USA
}
\affiliation{Department of Astrophysics,
 American Museum of Natural History,
Central Park West 79 St., NY 10024, USA
}


\date{Aug 28th 2015}
\begin{abstract}
  \noindent The sharp change in slope of the ultrahigh energy cosmic
  ray (UHECR) spectrum around \energy{18.6} (the ankle), combined with
  evidence of a light but extragalactic component near and below the
  ankle and intermediate composition above, has proved exceedingly
  challenging to understand theoretically, without fine-tuning.  We propose a mechanism
  whereby photo-disintegration of ultrahigh energy nuclei in the
  region surrounding a UHECR accelerator
 accounts for the observed spectrum and inferred
  composition at Earth.  For suitable source
  conditions, the model reproduces the spectrum and the composition
  over the entire extragalactic cosmic ray energy range, i.e. above
  \energy{17.5}.  Predictions for the spectrum and flavors of
  neutrinos resulting from this process are also presented.
\end{abstract}

\maketitle

\section{Introduction}

The cosmic ray spectrum spans roughly eleven decades of energy,
$\energy{9} \alt E \alt \energy{20}$ and has three major features: the
steepening of the spectrum dubbed the ``knee'' at
$\approx$\energy{15.6} ~\cite{Antoni:2005wq}, a pronounced hardening
of the spectrum at $E \approx \energy{18.6}$, the so-called ``ankle''
feature~\cite{Bird:1993yi, Abbasi:2007sv,Abraham:2010mj}, and finally a cutoff
around $10^{19.6}~{\rm eV}$~\cite{Abbasi:2007sv,Abraham:2008ru}.
Three additional more subtle features have been reported between the
knee and the ankle: A hardening of the spectrum at around
$2\times\energy{16}$~\cite{Apel:2012tda,Aartsen:2013wda,
  Knurenko:2013dia, Prosin:2014dxa} followed by two softenings at
$\sim \energy{16.9}$~\cite{Apel:2012tda, Aartsen:2013wda} and and
\energy{17.5}~\cite{Bird:1993yi, AbuZayyad:2000ay, Bergman:2007kn,
  Knurenko:2013dia, Prosin:2014dxa}. The latter is traditionally
referred to as the ``second knee''.

The variations of the spectral index reflect various aspects of cosmic
ray production, source distribution and propagation.  The first and
second knee have straightforward explanations, as reflecting the
maximum energy of Galactic magnetic confinement or acceleration
capability of the sources, both of which grow linearly in the charge
$Z$ of the nucleus; the first knee being where protons drop out and
the second knee where the highest-$Z$ Galactic cosmic rays drop out.
As the energy increases above the second knee to the ankle, the
composition evolves from heavy to light~\cite{Kampert:2012mx} while
the cosmic ray arrival directions are isotropic to high accuracy
throughout the
range~\cite{Abreu:2011ve,Auger:2012an,ThePierreAuger:2014nja}.
Finally, as the energy increases above the ankle, not only does the
spectrum harden significantly, but the composition gradually becomes
heavier (interpreting the data using conventional extrapolations of
accelerator-constrained particle physics
models)~\cite{Aab:2014kda,Aab:2014aea}.

This observed evolution in the extragalactic cosmic ray composition
and spectral index presents a major conundrum.  A pure proton
composition might be compatible with the observed spectrum of
extragalactic cosmic rays~\cite{Berezinsky:2002nc} when allowance is
made for experimental uncertainties in the energy scale and the fact
that the real local source distribution is not homogeneous and
continuous~\cite{Ahlers:2012az} (although the sharpness of the ankle
is difficult to accommodate), but a pure proton composition is
incompatible with the depth-of-shower-maximum ($X_{\rm max}$)
distributions observed by Auger~\cite{Aab:2014kda,Aab:2014aea} unless
current extrapolations of particle physics are incorrect.  Moreover, a
fit of the spectrum with a pure proton composition seems to require a
very strong source evolution~\cite{Fukushima:2015bza} which leads
to a predicted neutrino flux in excess of experimental limits~\cite{Aloisio:2015ega}.
 On the
other hand, models which fit the spectrum and composition at highest
energies, predict a deep gap between the end of the Galactic cosmic
rays and the onset of the extragalactic cosmic
rays~\cite{Allard:2005cx,Allard:2007gx,Allard:2008gj,DeDonato:2008wq,Taylor:2013gga,Deligny:2014opa}. Models
can be devised to fill this gap, but fine-tuning is required to
position this new population so as to just fit and fill the
gap~\cite{Gaisser:2013bla,Aloisio:2013hya,Giacinti:2015hva}.

\begin{figure*}[t!]
\centering \includegraphics[width=0.7\linewidth]{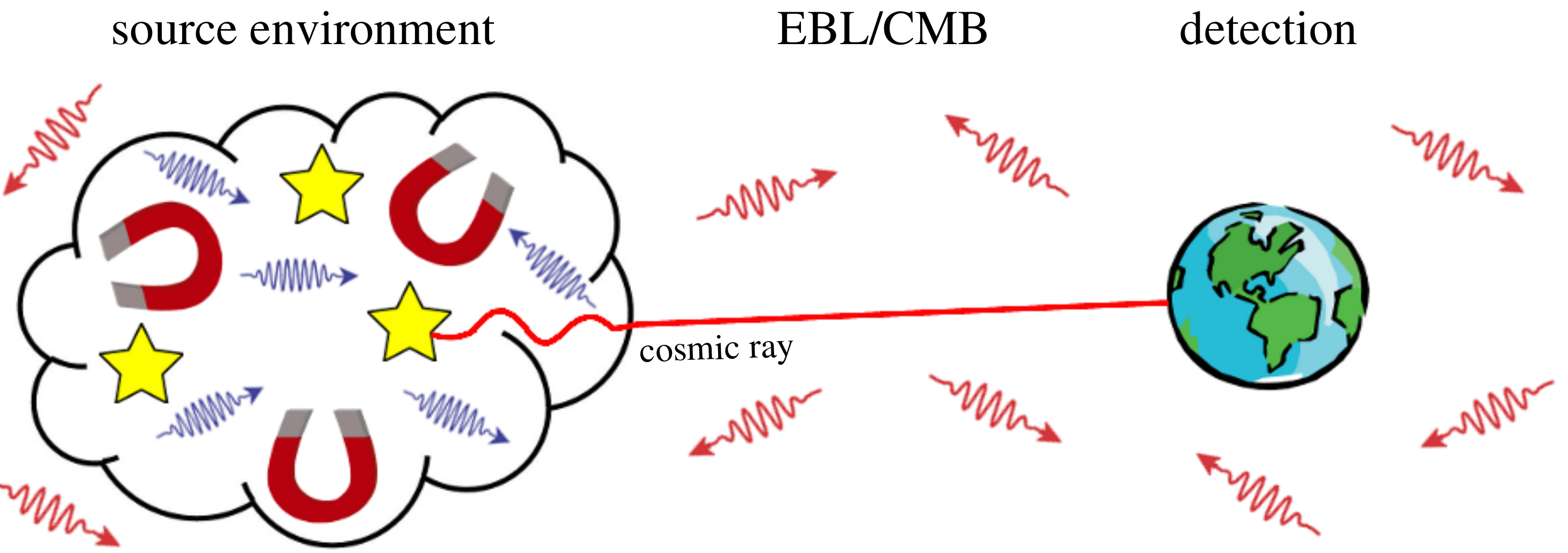}
\caption[illustration]{Illustration of our model calculation: Sources
  (yellow stars) inject cosmic rays with a power law in
  energy, into a surrounding region of radiation and turbulent magnetic fields.  After propagation through this local environment and then intergalactic space, these cosmic rays and their spallation products
  are detected at Earth.  The photon energies in the source environment are characteristically of much higher energy than in the extragalactic background light.}
\label{fig:illustration}
\end{figure*}

Here we offer a resolution to this conundrum, by showing that
``post-processing'' of UHECRs via
photo-disintegration in the environment surrounding the source, can
naturally explain the entire spectrum and composition.  In our model,
extragalactic cosmic rays below the ankle are predominantly protons
from nucleons knocked off higher energy nuclei in the region surrounding the
accelerator, and the spectrum and composition above the ankle are
predominantly dictated by the accelerator and propagation to Earth.
The model makes distinctive predictions about the spectrum and flavor
ratios of neutrinos, which should enable it to be tested.  If the
ankle and the protons below it arise on account of our mechanism, we
obtain a new constraint on UHECR sources beyond the Hillas criterion
and total-energy-injection requirements, namely that the environment
around the source has the conditions giving rise to the required
amount of photo-disintegration.

Up until now, photo-disintegration (PD) has been mainly considered as
a danger inside the accelerator, as it would cut off the cosmic ray
spectrum at energies such that the PD interaction length and the
acceleration length are comparable.  Since the acceleration length
increases with energy, whereas the PD interaction length generally
decreases with energy, photo-dissociation acts as a {\itshape low-pass
  filter}.  The insight underlying the mechanism we propose, is that
if the primary locus of PD is {\itshape outside} the accelerator, PD
generally acts as a {\itshape high-pass filter}, permitting the
highest energy cosmic rays to escape unscathed while the lower energy
ones are disintegrated inside the source region, generating nucleons
with energy $1/A$ of the original nucleus of mass $A$.  As we shall
see, these spallated nucleons naturally produce the ankle feature,
explain why extragalactic cosmic rays below the ankle are protonic,
and account for the spectral index below the ankle.  Examples of
systems in which the accelerator is embedded in a photon field and the
cosmic rays are trapped by magnetic fields in that environment could
be the dusty torus surrounding an active galactic nucleus or the interstellar medium of the star-forming region surrounding most young
pulsars; see
also~\cite{Allard:2009fb,Kotera:2009ms,Pe'er:2009rc,Fang:2012rx,Fang:2013cba,Globus:2014fka,Parizot:2014ixa,Kotera:2015pya}.
The basic setup of our phenomenological model is illustrated in
Fig.\,\ref{fig:illustration}.

The layout of the paper is as follows. In Sec.\,\ref{sec:ankle} we
introduce our model and in Sec.\,\ref{sec:datacomparison} we compare
its predictions with experimental data. Details about particle
propagation and the calculation of multi-messenger signatures are
given in the appendices. Section~\ref{sec:conclusions} contains our
conclusions.

\section{Formation of the Ankle}
\label{sec:ankle}

To illustrate the mechanism we have identified to create the ankle and
generate protons below, consider a system in which the accelerator
(also referred to as the source) is embedded in an environment in which the
cosmic rays 
are confined for some time by magnetic fields
while interacting with the ambient radiation field.
Our essential simplifications are:
{\it (i)} a fast acceleration mechanism and/or a low photon density
inside the accelerator,
{\it (ii)}~no energy is lost except through an interaction, and
whenever a nucleus interacts it loses one or more nucleons by
photo-disintegration or photo-pion production (in this case the
nucleus loses a fraction of its energy corresponding to the reduction
in its nuclear mass);
{\it (iii)}~a cosmic ray either escapes without changing energy, with a
rate $\tau_\mathrm{esc} $, or the cosmic ray interacts one or more
times before escaping;
{\it (iv)}~$\tau_\mathrm{esc} $ and $\tau_\mathrm{int} $ are
independent of position in the source environment and depend only on
$\{E, A, Z\}$ of the nucleus.  In this approximation the number of
nuclei in a given energy range and with a specified $\{A, Z\}$
decreases exponentially with time, with

\begin{equation}
\tau =
(\tau_\mathrm{esc}^{-1} + \tau_\mathrm{int}^{-1})^{-1} \, .
\end{equation}
  A fraction
\begin{equation}
\eta_\mathrm{esc} = (1 + \tau_\mathrm{esc}/\tau_\mathrm{int})^{-1}
\label{eta1}
\end{equation}
of the particles escape without interaction and the rest interact
before escaping, so $\eta_\mathrm{int} = 1 - \eta_\mathrm{esc}$.  Note
that $\eta_\mathrm{esc}$ and $\eta_\mathrm{int}$ depend only on the
ratio of the escape and interaction times, but not on the absolute
value of either of them.

A simple analytic treatment is instructive.  To illustrate the
low/high-pass filter mechanism, consider the case that the escape and
interaction times are both power laws in energy,
\begin{equation}
\tau_\mathrm{esc} = a\,(E/E_0)^\delta \quad {\rm and} \quad
\tau_\mathrm{int} = b\,(E/E_0)^\zeta.
\label{taus}
\end{equation}
Then
\begin{equation}
\eta_\mathrm{esc}(E) = \left(1 +
  R_0\,(E/E_0)^{\delta-\zeta}\right)^{-1} \,,
\end{equation}
where $R_0=a/b$ is the ratio of the escape and interaction time at
reference energy $E_0$.  When $\delta>\zeta$, the source environment
acts as a {\itshape low-pass filter} on the particles injected from
the accelerator, leading to a cutoff in the escaping spectrum at high
energies. This situation is typical of leaky box models of
diffuse acceleration at time-independent shocks~\cite{Szabo:1994qx, Protheroe:1998pj, Drury:1999ri} where
$\delta>0$ because the higher the energy of the particle, the longer
it needs to stay in the accelerator to reach its energy.
 By contrast, if the
escape time decreases with energy, as in the case of diffusion in
turbulent magnetic fields outside the accelerator, then it is possible
to have $\delta<\zeta$ leading to a {\itshape high-pass filter} on the
energy spectrum of injected nuclei: the lower the energy, the more
time the nuclei have to interact before escaping, leading to a
hardening of the spectrum and lightening of the composition of nuclei
escaping the region surrounding the source.  The spallated nucleons
have energies of $E = E_{A} / A$; these nucleons are most
abundant at low energies and have a steeper spectrum $\varpropto (1 -
\eta_\mathrm{esc}(E^* \, A^\prime))$.  Thus the high-pass scenario
leads naturally to an ankle-like feature separating the nucleonic
fragments from the remaining nuclei.  The normalization and slope of
the spectrum of spallated nucleons relative to that of the primary
nuclei is determined by how thoroughly the primary nuclei are
disintegrated, which is governed by the ratio of escape and
interaction lengths of the most abundant primaries.

\begin{figure}[t!]
   \centering \includegraphics[clip,
     width=\linewidth]{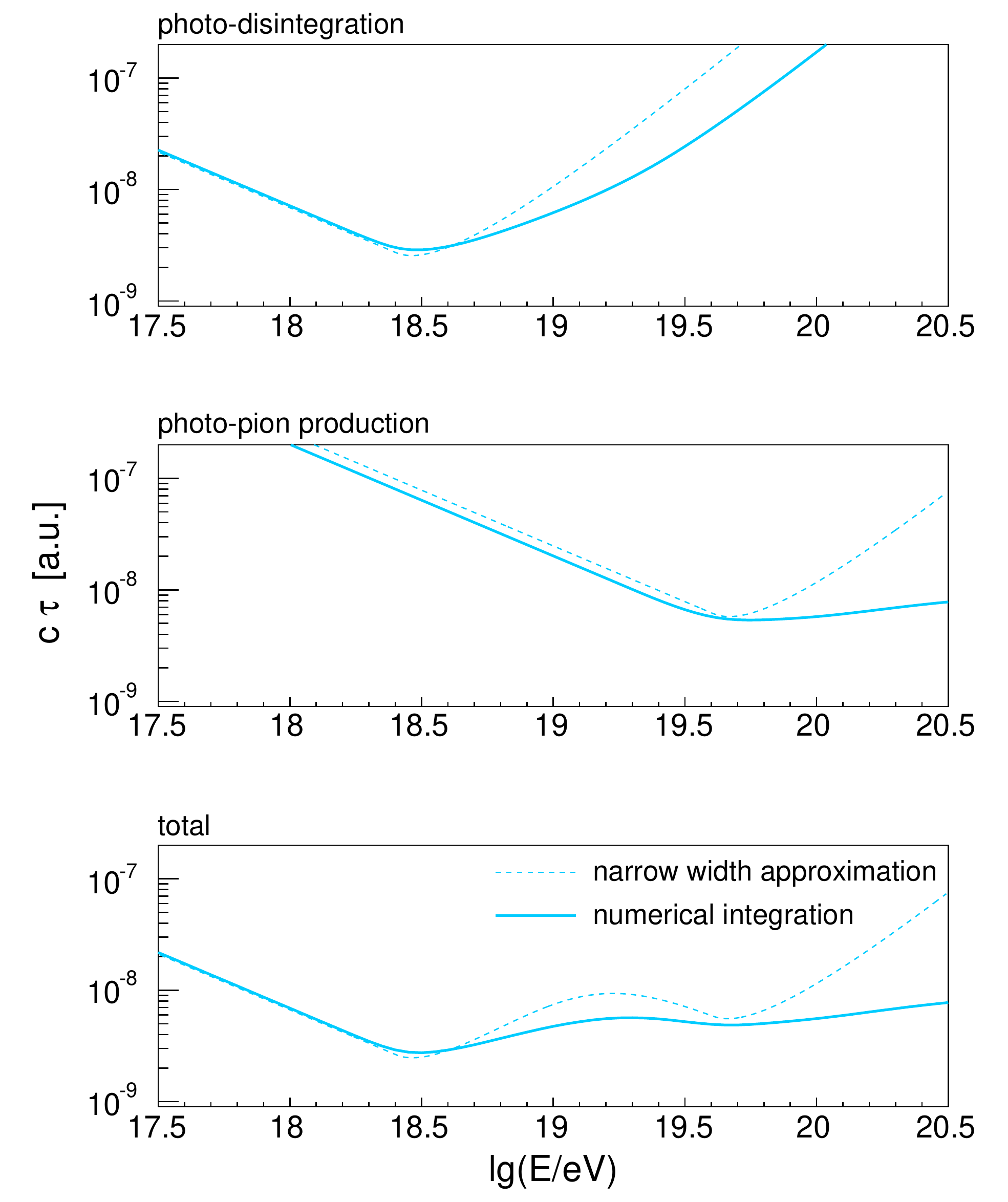} \caption[tauInt]{Interaction
     times of $^{28}$Si in a broken power-law photon field with
     parameters $\alpha=\frac{3}{2}$, $\beta=-1$ and $\varepsilon_0 =
     0.11~{\rm eV}$. Top panel: photo-disintegration, middle panel:
     photo-pion production, bottom panel: sum of the two
     processes. The results of numerical integration using
     detailed cross sections are shown as thick solid lines, while those of the
     narrow-resonance-approximation (detailed in Appendix~\ref{app:appendixB}) are displayed
     with thin dashed lines.}
\label{fig:vwl}
\end{figure}

To obtain a more realistic treatment of the interaction time, we must
specify the shape of the spectrum of the target photons. In our work
to date we have considered: {\it (i)}~a broken power-law (BPL),
characterized by its peak energy $\epsilon_{0}$ and lower and upper
spectral indices $\alpha,\beta$ (this is a simplified representative
of non-thermal emission that allows for analytic calculation as discussed below and in Appendix~\ref{app:appendixB}); {\it (ii)} a black-body spectrum; {\it
  (iii)} two types of modified black-body spectrum, which result from
a reprocessed black-body in a dusty
environment~\cite{kruegel}. Details are given in
Appendix~\ref{app:appendixA}.  For such peaky photon spectra the
interaction time does not have the simple representation of
(\ref{taus}) but it does have a rather universal structure.   In our actual calculations we adopt a numerical integration of
{\sc Talys}~\cite{talys, TALYSRestored}
\nocite{Batista:2015mea} 
and {\sc Sophia}~\cite{Mucke:1999yb}
cross sections using~\cite{crpropaTools}, but the analytic expression for $\tau_{\rm int}$ derived in
Appendix~\ref{app:appendixB} for the BPL in the narrow-width approximation for the interaction cross sections, is qualitatively similar and useful for understanding.  As can be seen in Fig.\,\ref{fig:vwl} the
folding of a single resonance with a broken power-law spectrum leads
to a ``V'' shape curve for $\tau_{\rm int}$ in a log-log plot for both photo-disintegration (top panel) and photopion production (middle panel).  Combining both processes in narrow-resonance approximation yields an interaction time with a
``W'' shape, while numerical integration including the plateau
for multi-pion production softens the ``W'' to what we shall refer to as an ``L'' shape for brevity, a shown in the bottom panel of Fig.\,\ref{fig:vwl}.  As evident from Fig.\,\ref{fig:vwl}, below the inflection point for photodisintegration $E_{b}$, the narrow-resonance approximation is good, while from the full numerical integration in the high-energy region $\tau_{\rm int}$ is roughly constant, so using the BPL spectrum, we have the approximate representation:
\begin{equation}
\tau_{\rm int} (E) \approx \tau_b \left\{ \begin{array}{lr}
  (E/E_b)^{\beta+1} & E \leq E_b \\ 1 & E> E_b \\
\end{array} \right. \,,
\label{tautau}
\end{equation}
where formulae for $\tau_b$ and $E_b$ are given in
Appendix~\ref{app:appendixB}, and the parameter values for photodisintegration are to be used.

Returning to the discussion  of $\tau_{\rm int} $ in (\ref{taus}) with (\ref{tautau})
yields the fraction of nuclei which escape without interaction in a
peaky photon spectrum. It is straightforward to see that if $\delta
<0$ and the interaction time is described by an L-shaped curve, then
$\eta_{\rm esc}$ has the properties of a high-pass filter.  These
conclusions do not depend on the exact shape of the photon
spectrum. As can be seen in Fig.\,\ref{app:photon-spectra} of
Appendix~\ref{app:appendixA}, the interaction times flatten to an
L-curve as well if the photon density is assumed to follow a (modified
) black body spectrum.

\section{Comparison with experiment}
\label{sec:datacomparison}

\subsection{Fiducial Model}
\label{sec:fidmodel}

As our fiducial example, we adopt a broken power-law photon spectrum
as a simplified representative of non-thermal emission given by
\begin{equation}
      n(\varepsilon) = n_0^{\rm BPL}
        \begin{cases}
           (\varepsilon/\varepsilon_0)^\alpha & \varepsilon < \varepsilon_0 \\
           (\varepsilon/\varepsilon_0)^\beta & \text{otherwise} \, .
        \end{cases}
\label{main:eq:photonfield}
\end{equation}
where $\varepsilon$ is the photon energy, the maximum photon number density is at
an energy of $\varepsilon_0$ and following~\cite{Szabo:1994qx} we take the slope parameters $\alpha =
+\frac{3}{2}$ and $\beta = -2$.  As we shall see later, any peaky spectrum gives similar results, with the position of the peak, $\varepsilon_0$, being the most important parameter besides the peak photon density.

Inspired by the energy dependence of
the diffusion coefficient for propagation in a turbulent magnetic
field, we model $\tau_\mathrm{esc}$ as a power law in rigidity $E/Z$,
\begin{equation}
\tau_\mathrm{esc} = \tau_0 (E Z^{-1}/E_0)^\delta.
\end{equation}
Since only the
ratio of escape and interaction times matters, and the $\{E, A, Z\}$
dependence of this ratio is entirely determined once the spectral
index of the escape time $\delta$ is specified, the remaining freedom
in characterizing the source environment can be encoded by specifying
the ratio of escape to interaction time for a particular choice of
$\{E, A, Z\}$, which we take at \energy{19} for iron nuclei, denoted
$R_{19}^{\rm Fe}$.  In application to a particular source candidate, $R_{19}^{\rm Fe}$ depends on the density of photons and the properties of the turbulent magnetic field that delays the escape of the UHECRs from the environment of their source.

\begin{figure}[t!]
   \centering
   \includegraphics[clip, viewport = 8 25 180 122,
   width=0.9\linewidth]{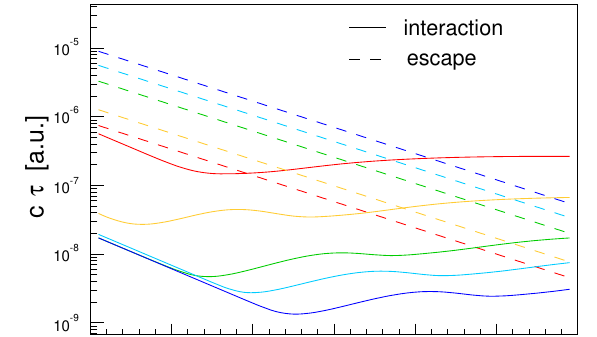}
    \includegraphics[clip, viewport = 197 140 369 258,
   width=0.9\linewidth]{PRDFiducial.pdf}
   \caption[source]{ {\bf Top:} Interaction and escape times for
     $A = 1,\, 4,\, 14,\, 28$ and $56$ (bottom to top for
     escape; vice versa for interaction) for the fiducial model photon field with $\varepsilon_0 = 0.11$ eV.  {\bf Bottom:} Injected
     $^{28}$Si flux (bold dashed) and escaping fluxes: thin black
     solid line denotes the sum of all escaping nuclei and solid
     curves give contribution of different mass groups with low energy
     intercept increasing with mass. Nucleons from photo-dissociation
     and photo-pion-production are shown with thin-dashed and dotted
     curves, respectively.}
\label{fig:tau&escspec}
\end{figure}

Figure~\ref{fig:tau&escspec}, upper panel, shows the escape and
interaction times in the fiducial source environment, as a function of
the cosmic ray energy, for proton, He, N, Si and Fe; the interaction
times are calculated including both photo-disintegration and
photo-pion production.  The gross features of the energy dependence of
the interaction times can be understood in the approximation of
resonant interactions in the nucleus rest frame
$\varepsilon'_\mathrm{res}$.  At low cosmic-ray energies, reaching
$\varepsilon'_\mathrm{res}$ requires high photon energy ($\varepsilon
> \varepsilon_0$), so that the interaction time decreases with
increasing cosmic-ray energy as $\tau \varpropto E^{\beta +1}$.  However for
high enough cosmic ray energy, the resonance can be reached in
collisions with photons of $\varepsilon < \varepsilon_0$. From here,
as the cosmic ray energy increases, the photon density at the resonant energy decreases as
$\varepsilon^{\alpha}$, and correspondingly the interaction times
increase. The laboratory energy of the inflection point of the
interaction times for a cosmic ray nucleus of mass $A m_p$ is at $E =
A m_p \varepsilon'_\mathrm{res} / (2 \varepsilon_0)$. The inflection
point of the photo-dissociation times can be seen as a dip in the plot
in the upper panel of Fig.\,\ref{fig:tau&escspec}, e.g., at around
\energy{18.8} for iron nuclei. At slightly higher energy, photo-pion
production becomes important, with the result that the energy
dependence of the interaction time is roughly speaking an L-shaped
curve in a log-log presentation.

Using these energy-dependent interaction and escape times, we
propagate nuclei through the source environment with the procedure
described in Appendix~\ref{app:appendixC}.  Cosmic rays of some given
composition are injected from the accelerator into the source
environment with a power law spectrum and an exponential cutoff at
some maximum rigidity.  To keep the complexity of the fiducial model to a
minimum, we inject only a single nuclear species and fix the
injection spectral index $\gamma = -1$, as expected for acceleration in young
neutron stars~\cite{Blasi:2000xm}.  The particles escaping the source
environment are then propagated through the intergalactic medium using
the procedure explained in Appendix~\ref{app:appendixD}.

In total the
fiducial model has 14 parameters, with 8
parameters allowed to float freely in the fit, as
indicated in Table~\ref{table:fiducial}.
The spectral index and normalization of the Galactic
spectrum are free ``nuisance'' parameters with the best fit giving a
spectral index of $-4.2$.  This should be understood as an effective
spectral index describing the cutoff of the Galactic cosmic ray
population, and hence cannot be directly compared with the parameter
reported by the KASCADE-Grande Collaboration~\cite{Apel:2013ura},
because their single-power law fit is driven by the ``low-energy''
data. The fraction of Galactic cosmic rays at \energy{17.5} is 55\%.

The best description of the data is obtained with
$^{28}$Si of maximum energy $Z \, \energy{18.5} = 4.6 \times
10^{19}$~\eV; the impact of allowing other parameters to vary is
discussed in following sections.   Normalizing this model to the observed flux at Earth, we
infer a comoving volumetric energy injection rate in CRs at $z=0$,
above \energy{17.5}, of $\bolddot{\epsilon}_{17.5} = 9.2 \times
10^{44}~{\rm erg \, Mpc^{-3}\, yr^{-1}}$.

The unmodified injection spectrum and the spectrum of escaping nuclei
for this fiducial model are shown in the lower panel of
Fig.\,\ref{fig:tau&escspec}.  At low energies, the nuclei are depleted
relative to the injected flux because $\tau_{\rm esc} \gg \tau_{\rm
  int}$, but the escaping nuclei follow the original spectral index
because in this example the interaction and escape times are parallel,
as to be expected for $\delta = \beta + 1$.  Once the corner of the
L-shape is reached, the fraction of escaping nuclei grows, leading to
an apparent hardening of the spectral index.

\begin{figure*}[th!]
\begin{center}
\subfigure[\enskip Flux at Earth]{
  \includegraphics[clip, viewport = 385 145 559 270, width=0.48\linewidth]{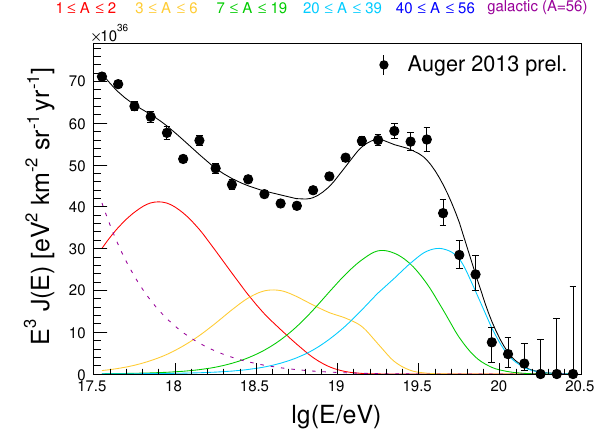}
\label{fig:flux}
}
\subfigure[\enskip Composition at Earth]{
  \includegraphics[clip, viewport = 385 8 559 133, width=0.48\linewidth]{PRDFiducial_fig4.pdf}
\label{fig:composition}
}
\end{center}
\caption[earth]{Spectrum and composition at Earth. The data points are from
  the Pierre Auger Observatory~\cite{ThePierreAuger:2013eja,
    Aab:2014kda}, error bars denote the statistical uncertainties and
  the shaded boxes illustrate the experimental systematic
  uncertainties of the composition. The composition estimates are
  based on an interpretation of air shower data with \Epos; the lines
  denote the predictions of our fiducial model.}
\label{fig:fit}
\end{figure*}

Even for the simple case in which a single nuclear species
is injected into the source environment, we obtain a complex evolution of the mass
composition with energy. At low energies the composition is dominated
by knock-off nucleons whereas at high energies the composition becomes
heavier as the ratio of escape to interaction time drops and more
heavy nuclei can escape before interacting.

This fiducial model of interactions in the source environment is a very simple one, yet even so it offers a
remarkably good accounting for the flux and composition at Earth as
determined by the Pierre Auger Observatory.  (Data from
the Telescope Array (TA) are consistent with the Auger results within
systematic and statistical uncertainties~\cite{Dawson:2013wsa,
  Abbasi:2015xga} and also can be well-fit; we come to TA separately below.)  In
Fig.\,\ref{fig:fit} we compare the fiducial model prediction to the Auger measured flux, from
\energy{17.5} to above \energy{20}~\cite{ThePierreAuger:2013eja} and to
the mean and variance of the distribution of the logarithm of mass on
top of the atmosphere, $\langle \ln A \rangle$ and $V(\ln
A)$~\cite{Abreu:2013env, AhnICRC13, Aab:2014kda}.
\begin{figure*}[hbt!]
\includegraphics[clip, viewport = 9 390 490 739, width=0.49\linewidth]{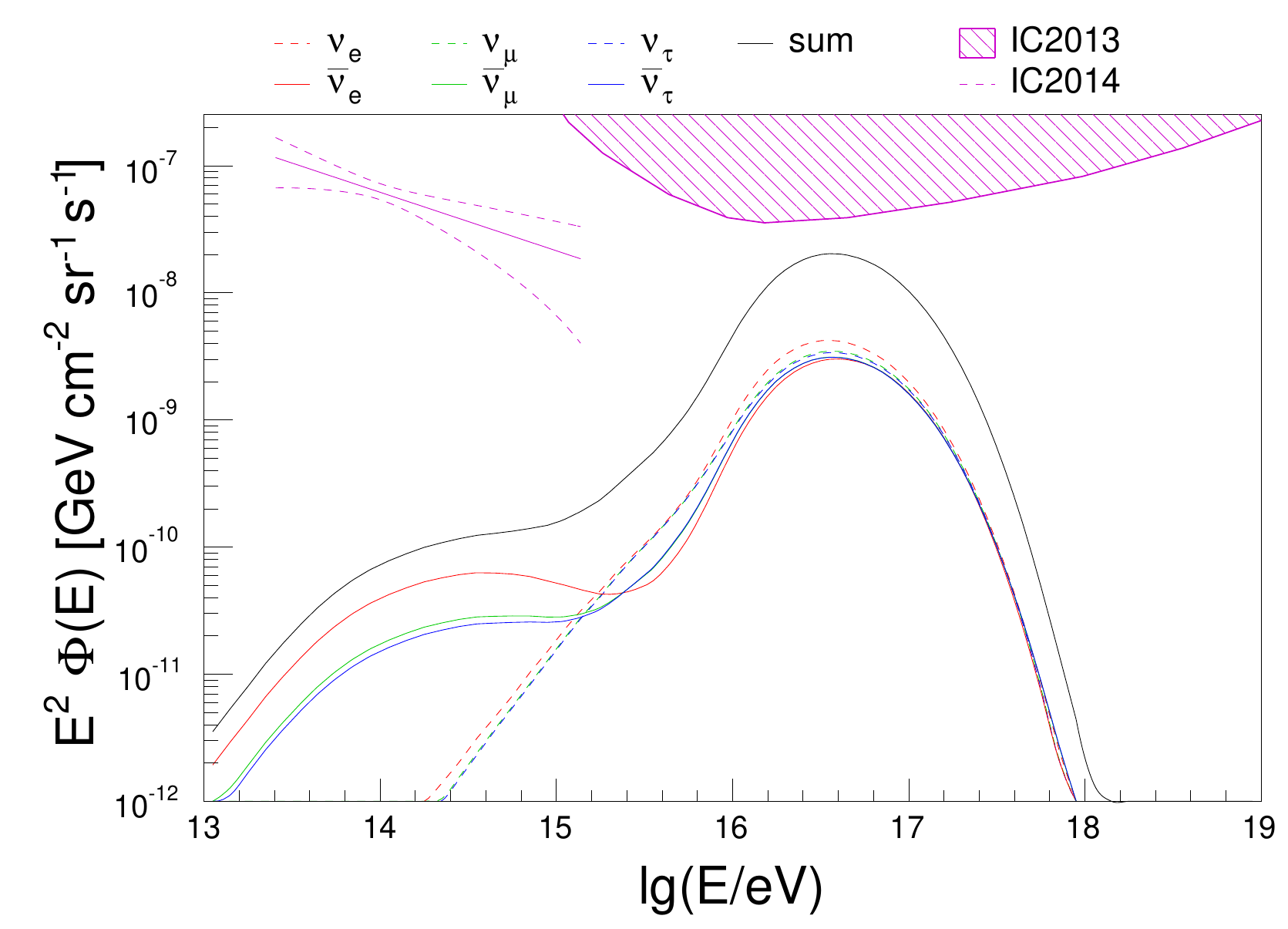}\quad
\includegraphics[clip, viewport = 9 20 490 369, width=0.49\linewidth]{PRDFiducial_nu.pdf}
\caption[earth]{Neutrino spectrum (left) and expected number of events in 10 IC86-years (right) for the fiducial model.
The measured flux of low-energy extragalactic neutrinos from
IceCube~\cite{Aartsen:2014muf} is shown in the left panel (purple
lines) as well as the 90\% CL upper limit on the flux of high-energy
neutrinos (dashed area)~\cite{Aartsen:2013dsm}. The peak in the
electron neutrino flux at about $10^{15.8}$ eV seen in the right panel is due to the increased
interaction probability of anti-electron neutrinos at the Glashow
resonance.
\label{fig:neutrinos}}
\end{figure*}
There is a good overall agreement between the model and the data. The
shape of the spectrum is described well, including the ankle and the
flux suppression.  The model also qualitatively reproduces the increase
of the average logarithmic mass with energy and the decrease of its
variance.

The neutrino signals of the fiducial model are shown in
Fig.\,\ref{fig:neutrinos}; details of the calculation are given in
Appendix ~\ref{app:appendixE}.  An exciting aspect of our model for the ankle is the presence of a detectable anti-electron-neutrino flux from neutron $\beta$-decay, with a rate consistent 
with the na\"{\i}ve estimate
of~\cite{Anchordoqui:2014pca}.  The right panel of Fig.\,\ref{fig:neutrinos}
shows the number of events as a function of energy predicted in ten
years of IC86, using the IceCube acceptance for different neutrino
flavors given in~\cite{Aartsen:2013dsm}.  In total, the fiducial model predicts
3.5 events in the range $10^{16}-10^{17}$~\eV after 10 years of
operation of IceCube (corresponding to about one year of operation for an upgraded
IceCube-Gen2 detector~\cite{Aartsen:2014njl}).  We emphasize the
distinctive $\bar{\nu}_{e}$ enrichment due to beta decay of spallated
neutrons.

The associated photon flux from nuclear de-excitation in our model is well below the Fermi-LAT data (see Appendix ~\ref{app:appendixE} for more detail).
Photo-pion interactions at the source and during propagation produce
an additional flux of photons via $\pi^0$-decay;  this is consistent with Fermi-LAT data, as follows:  If the
origin of the photons measured by Fermi-LAT is exclusively from these
interactions, then from ~\cite{Ahlers:2010fw} the associated diffuse neutrino
flux saturates the IceCube upper
limit~\cite{Aartsen:2013dsm}. Since the neutrino flux in the fiducial
model is below the IceCube limit, it follows that also the associated
photon flux is consistent with Fermi-LAT data.  A more sophisticated realization of our mechanism than in the fiducial model must also respect the IceCube limits, and therefore the Fermi-LAT data as well.

\begin{table*}[ht]
\begin{tabular}{l c l l }
\hline
\hline
{\itshape source parameters} \\
\hline
power law index of injected nuclei & $\gamma$ & fix\quad\quad \quad & $-1$ \\
mass number of injected nuclei & $A$& free & 28 (29)  \\
maximum energy & $E^p_{\rm max}$ & free  & \energy{18.5\;(18.6)} \\
cosmic ray power density, $E>\energy{17.5}$ &
$\bolddot{\epsilon}_{17.5}$  & free & 9.2 (13) $\times 10^{44}~{\rm erg \,  Mpc^{-3}\, yr^{-1}}$ \\
evolution & $\xi(z(t))$ & fix & star formation rate~\cite{Robertson:2015uda} \\ [1ex]
\hline
{\itshape source environment} \\
\hline
energy of maximum of photon field density & $\varepsilon_0$ &free &
$0.11\;(0.07)~{\rm eV}$  \\
power law index of photon spectrum ($\varepsilon < \varepsilon_0$) & $\alpha$ & fix & $+\frac{3}{2}$ \\
power law index of photon spectrum  ($\varepsilon \geq \varepsilon_0$) & $\beta$ & fix & $-2$\\
power law index of escape length & $\delta$ & free & $-0.77$ ($-0.94$) \\
ratio of interaction and escape time & $R_{19}^{\rm Fe}$ & free
& 4.4 (3.7) $\times 10^2$ \\ [1ex]
\hline
{\itshape propagation to Earth} \\
\hline
infra-red photon background & -- & fix &
Gilmore12~\cite{Gilmore:2011ks}  \\ [1ex]
\hline
{\itshape spectrum of Galactic cosmic rays} \\
\hline
power law index at Earth& $\gamma_{\rm gal}$ & free & $-4.2$ ($-3.7$)  \\
mass number of Galactic nuclei & $A_{\rm gal}$& fix & 56 \\
flux fraction at \energy{17.5} & $f_{\rm gal}$& free & 57 (72) \% \\
\hline
\hline
\end{tabular}
\caption[parameters]{Parameters of the fiducial model. Values in
  brackets denote the parameters of the best-fit obtained when shifting the data by
  its systematic uncertainties (see text).}
\label{table:fiducial}
\end{table*}

\subsection{Model Variations}
\label{sec:modelvariations}

In this section we discuss the impact of theoretical and experimental
uncertainties on our model, as well as different choices for the
fiducial parameters.

\subsubsection{Experimental Uncertainties}
To study the influence of the experimental systematic uncertainties on
our fit, we have repeated the fit for all combinations of altering the
measurements by $+1$, $+0$ and $-1\,\sigma_{\rm sys.}$ of the quoted
uncertainties on the energy and composition scale.  We find that the
best fit is obtained within the experimental systematics when shifting
the energy scale up by $+1\,\sigma_{\rm sys.}  = +15\%$ and by
shifting $\langle \ln A \rangle$ and $V(\ln A)$ corresponding to a
shift of the shower maximum by $-1\,\sigma_{\rm sys.} \approx
-10~\gcm$.  The best-fit values after the application of these shifts
are shown in brackets in Table~\ref{table:fiducial}. Most notably, the
peak energy of the photon spectrum decreases from 110 to 70~\meV and
the best-fit value of the spectral index of the escape time decreases
from $\sim -3/4$ to almost $-1$. The neutrino flux at Earth obtained
for this fit is about 30\% smaller than in case of the fiducial
model. This is mainly due to the difference in the best-fit peak
energy of the photon field in the source environment.  The sensitivity
of the neutrino flux to $\varepsilon_0$ will be further discussed in
Sec.\,\ref{sec:modelVarPhotonSpec}.

\begin{figure*}[!t]
\subfigure[\enskip $\tau_{\rm int}$ and $\tau_{\rm esc}$ .]{
   \includegraphics[clip, viewport = 2 4 184 122,
   width=0.48\linewidth]{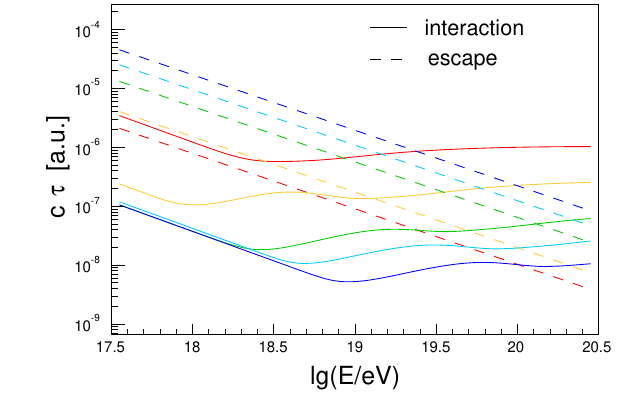}
   \label{fig:tauSys}
}
\subfigure[\enskip Injected (dashed line) and escaping (solid lines) fluxes.]{
   \includegraphics[clip, viewport = 197 140 379 258,
   width=0.48\linewidth]{PRDSys.pdf}
   \label{fig:escapeSys}
}\\
\subfigure[\enskip Flux at Earth]{
  \includegraphics[clip, viewport = 385 145 559 264, width=0.48\linewidth]{PRDSys.pdf}
\label{fig:fluxSys}
}
\subfigure[\enskip Composition at Earth]{
  \includegraphics[clip, viewport = 385 8 559 127, width=0.48\linewidth]{PRDSys.pdf}
\label{fig:compositionSys}
}
\caption[earth]{Spectrum and composition at Earth. The data points are from the
Pierre Auger Observatory~\cite{ThePierreAuger:2013eja, Aab:2014kda}
{\itshape shifted by plus one sigma of systematic uncertainty for
the energy scale and minus one sigma for the \Xmax scale.}
The lines denote the best-fit within our fiducial model.}
\label{fig:fitSys}
\end{figure*}

The overall description of the spectrum and composition is considerably
improved, as can be seen in Fig.\,\ref{fig:fitSys}. The model
variations discussed below will therefore be performed based on shifted data.

\subsubsection{Hadronic Interactions in Air Showers}
\label{sec:hadintatmo}
The interpretation of experimental air shower data in terms of mass
composition relies on the validity of extrapolations of the properties
of hadronic interactions to ultrahigh energies.  Using alternative
models for this interpretation (\Sibyll~\cite{Ahn:2009wx} or
\QgII~\cite{Ostapchenko:2010vb} instead of
\Epos~\cite{Pierog:2013ria}), decreases the value of the $\langle \ln
A \rangle$ data points by about $\langle \ln A \rangle = -0.6$ and
leads to a worse fit of the data.  If this difference between models
gives a fair estimate of the uncertainties of the mass determination
in both directions, $\sigma_\mathrm{theo}(\langle \ln A \rangle) = \pm
0.6$, then a hadronic interaction model that leads to a heavier
interpretation of Auger data than \Epos would make the fit with the
fiducial model even better, similar to the systematic shift in the
composition scale discussed in the previous section.

\subsubsection{Mass Composition at the Source}
\label{sec:masscompo}
It is remarkable that a good description of both the spectrum and
mass composition at Earth is possible by assuming only a single
injected species at the source as assumed for simplicity in the fiducial
model. However, depending on the astrophysical scenario, this might be
an unrealistic assumption.

\begin{figure*}[t!]
\begin{center}
\subfigure[\enskip Proton.]{
   \includegraphics[width=0.315\linewidth]{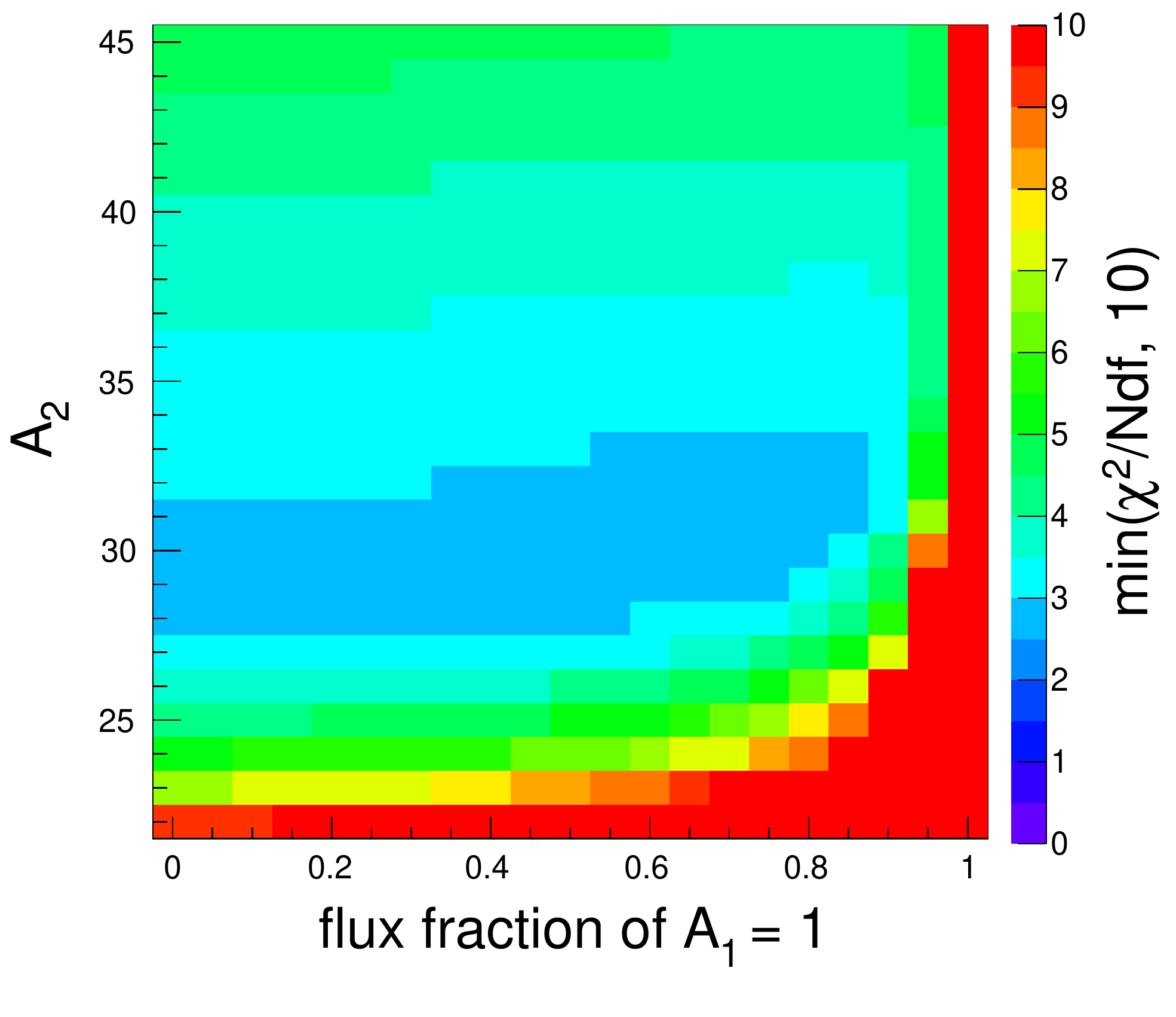}
   \label{fig:mass1}
}
\subfigure[\enskip Helium.]{
   \includegraphics[width=0.315\linewidth]{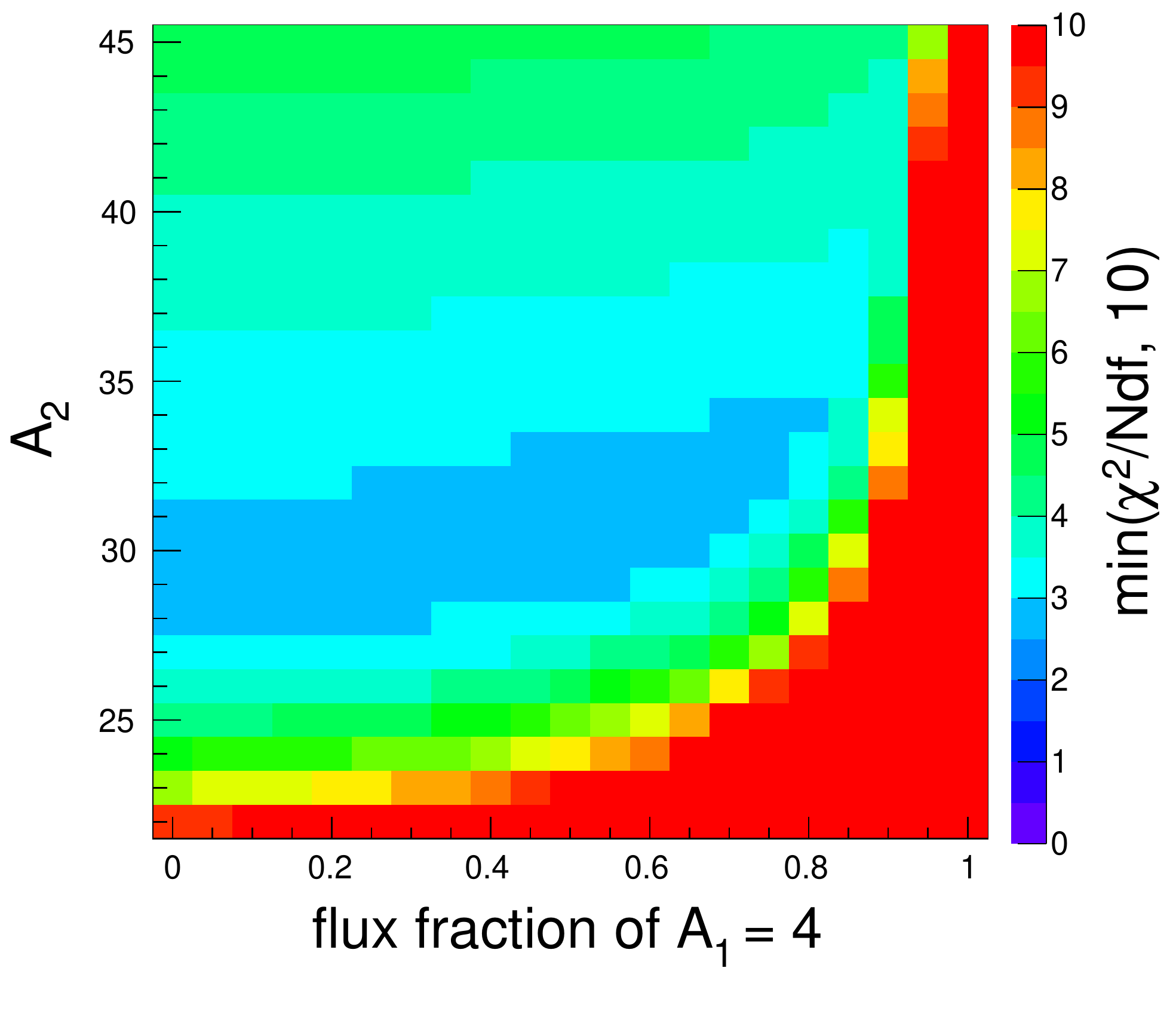}
   \label{fig:mass4}
}
\subfigure[\enskip Nitrogen.]{
   \includegraphics[width=0.315\linewidth]{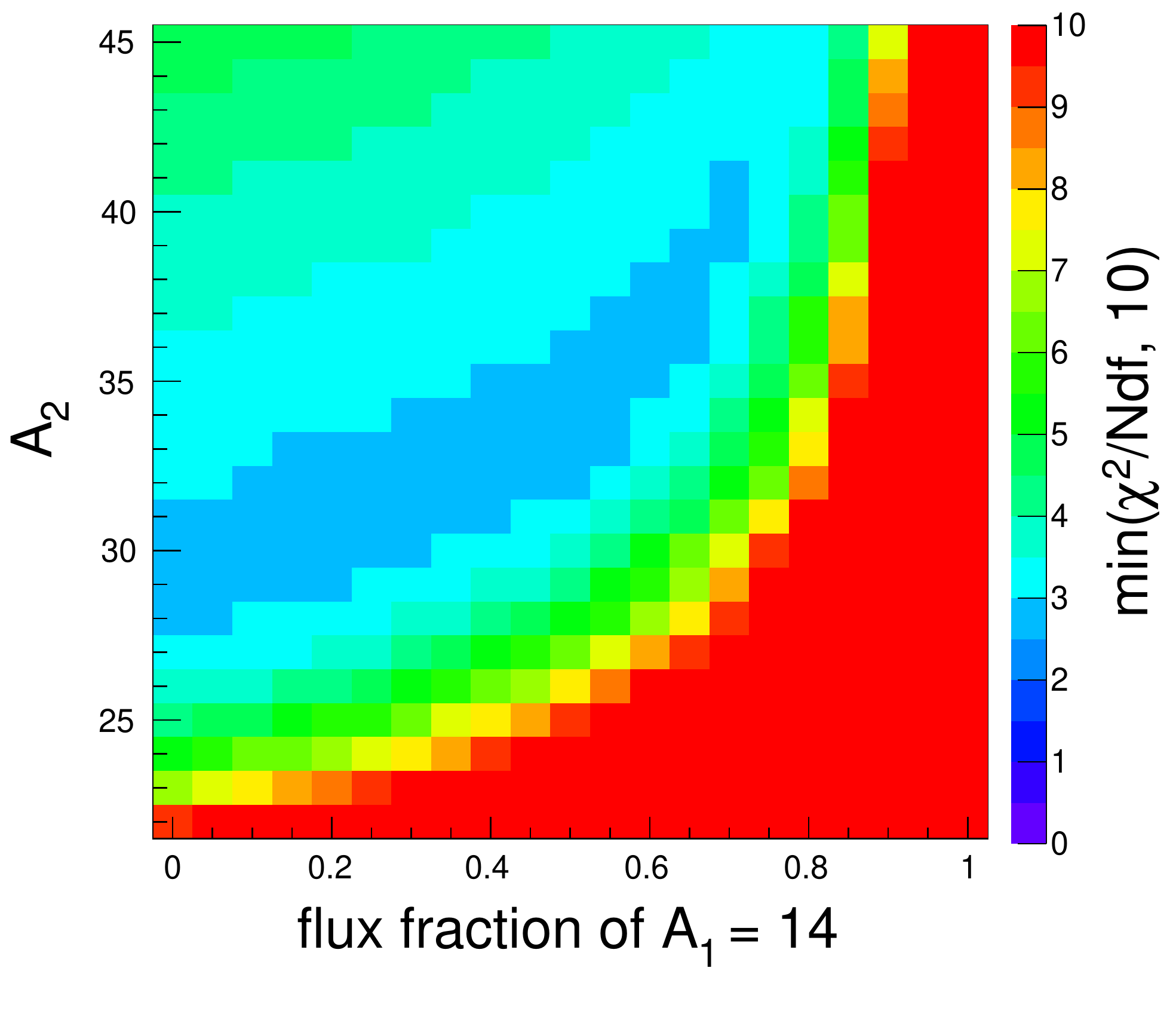}
   \label{fig:mass14}
}\\
\subfigure[\enskip Iron, Auger flux.]{
   \includegraphics[width=0.35\linewidth]{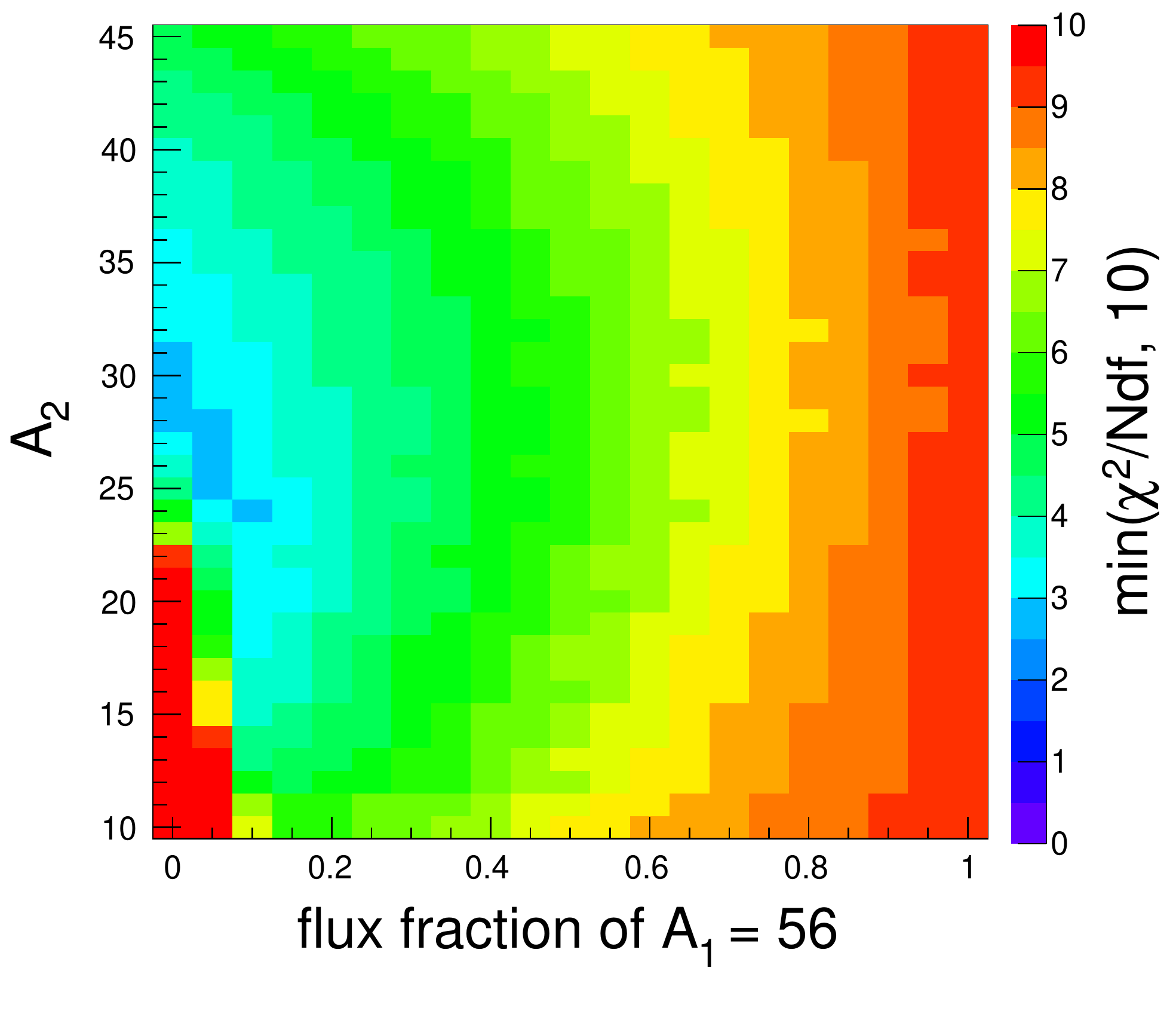}
   \label{fig:mass56}
}\quad
\subfigure[\enskip Iron, TA flux.]{
   \includegraphics[width=0.35\linewidth]{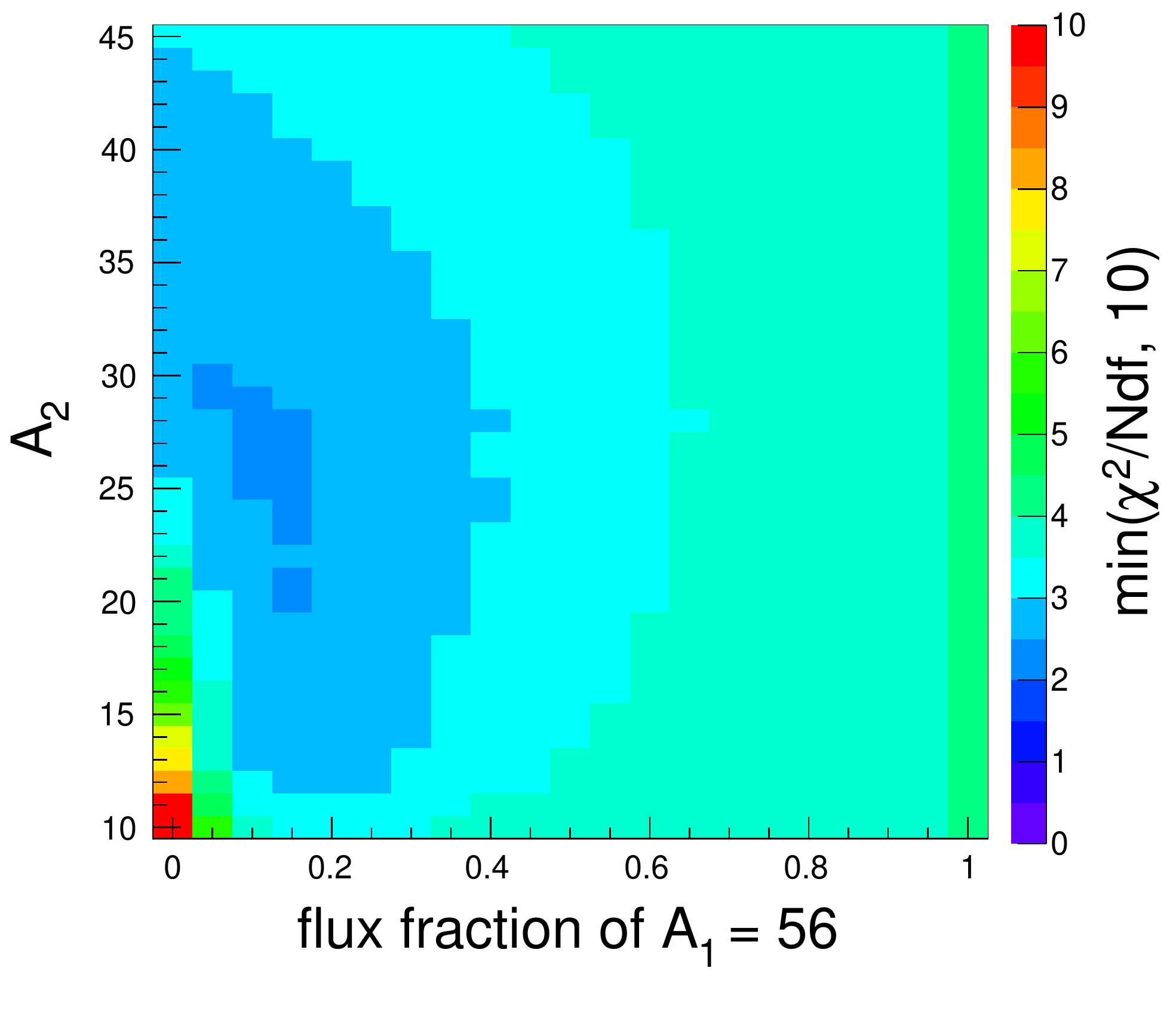}
   \label{fig:mass56TA}
}
\end{center}
\caption[mass scan]{Injection of two mass components. The first mass
  value, $A_1$, is fixed and contributes the fraction indicated on the
  x-axis to the total flux. The second mass value, $A_2$, is varied as
  shown on the y-axis. The fit quality is indicated by the colors.}
\label{fig:massScan}
\end{figure*}

In Fig.\,\ref{fig:massScan} we explore the capability of our model to
incorporate additional flux components of mass $A_1$ below and above
the mass $A_2 \sim 29$ that gives the best fit for the fiducial
single-mass model.  As can be seen, our calculation allows for an
additional proton or helium component as large as 80\% and up to 70\%
for nitrogen.

For an additional flux component with a heavy mass, the model is more
restrictive as illustrated in the lower left panel of
Fig.\,\ref{fig:massScan} using $A_1 = 56$. In this case, the
description of the data considerably deteriorates for fractions above
10\%. The reason for this behavior is twofold. Firstly, the injection
of too much iron at the source leads to a too heavy composition at
Earth as compared to the estimates from the Pierre Auger
Observatory. Secondly, if the end of the cosmic ray spectrum is to be
described by the maximum rigidity of iron nuclei, then the energy of
secondary nucleons needed to populate the flux at and below the ankle
is too small to describe the data (the maximum energy of secondary
nucleons is $1/A$ of the maximum energy of nuclei).

\begin{figure*}[t!]
\subfigure[\enskip $\tau_{\rm int}$ and $\tau_{\rm esc}$ .]{
   \includegraphics[clip, viewport = 2 4 184 122,
   width=0.48\linewidth]{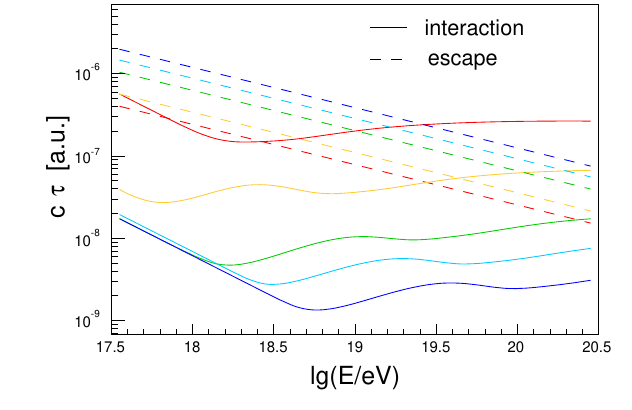}
   \label{fig:tauTA}
}
\subfigure[\enskip Injected (dashed line) and escaping (solid lines) fluxes.]{
   \includegraphics[clip, viewport = 197 140 379 258,
   width=0.48\linewidth]{PRDTA.pdf}
   \label{fig:escapeTA}
}\\
\subfigure[\enskip Flux at Earth]{
  \includegraphics[clip, viewport = 385 145 559 264, width=0.48\linewidth]{PRDTA.pdf}
\label{fig:fluxTA}
}
\subfigure[\enskip Composition at Earth]{
  \includegraphics[clip, viewport = 385 8 559 127, width=0.48\linewidth]{PRDTA.pdf}
\label{fig:compositionTA}
}
\caption[earth]{Spectrum and composition at Earth. The data points are from the
TA~\cite{AbuZayyad:2012ru} (flux) and the Pierre Auger
Observatory~\cite{Aab:2014kda} (composition). The latter have been
shifted in energy to match the energy scale of TA and
the \Xmax scale is shifted down by 1 sigma. The lines denote the fit
with our model assuming a pure iron composition at the source.}
\label{fig:fitTA}
\end{figure*}

If the cut-off of the flux is at higher energies, as suggested by the
measurement of TA~\cite{AbuZayyad:2012ru}, then a larger fraction of
iron primaries at the source can be incorporated, as shown in the
lower right panel of Fig.\,\ref{fig:massScan}.  When using the TA data
in the fit, as shown in Fig.\,\ref{fig:fitTA}, the spectrum can be
described reasonably well even for an injected flux consisting of
100\% iron nuclei. But in this scenario the composition at Earth at ultrahigh energies
is heavier than suggested by the interpretation of the $\Xmax$ data of
Auger.

\begin{figure*}[htb]
\begin{center}
\subfigure[\enskip Injected fluxes in 5 mass groups.]{
   \includegraphics[clip, viewport = 2 140 184 258,
   width=0.48\linewidth]{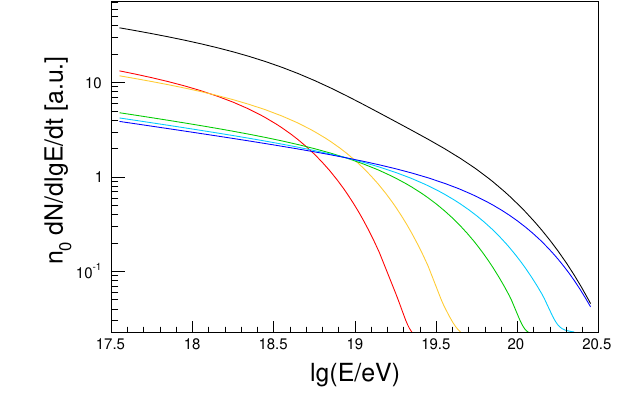}
   \label{fig:injectedGalactic}
}\subfigure[\enskip Escaping fluxes (sum of injection shown as dashed line).]{
   \includegraphics[clip, viewport = 197 140 379 258,
   width=0.48\linewidth]{PRDGalactic.pdf}
   \label{fig:escapeGalactic}
}\\
\subfigure[\enskip Flux at Earth]{
  \includegraphics[clip, viewport = 385 145 559 264, width=0.48\linewidth]{PRDGalactic.pdf}
\label{fig:fluxGalactic}
}
\subfigure[\enskip Composition at Earth]{
  \includegraphics[clip, viewport = 385 8 559 127, width=0.48\linewidth]{PRDGalactic.pdf}
\label{fig:compositionGalactic}
}
\end{center}
\caption[earth]{Spectrum and composition at Earth. The data points are from
  the Pierre Auger Observatory~\cite{ThePierreAuger:2013eja,
    Aab:2014kda} shifted by their systematic uncertainty as in
  Fig.\,\ref{fig:fitSys}. The injected composition follows a Galactic
  mixture with 10 elements (see text).}
\label{fig:fitGalactic}
\end{figure*}

As an illustration of a more complex composition model, we use the
abundances of Galactic nuclei at a nucleus energy of 1~\TeV, which we
read from Fig.\,28.1 in~\cite{Agashe:2014kda}. The flux fractions are
0.365, 0.309, 0.044, 0.077, 0.019, 0.039, 0.039, 0.0096, 0.014, 0.084
for H, He, C, O,Ne, Mg, Si, S, Ar+Ca, Fe, respectively. The resulting
fit is shown in Fig.\,\ref{fig:fitGalactic} ($\gamma = - 1.25$ and
$\delta=-1$). This example demonstrates that our mechanism for
producing the ankle is working even when considering a complicated mix
of primaries.

\subsubsection{Source Evolution and Spectral Index}
\label{sec:sourcevoAndGamma}
\begin{figure*}[p!]
\begin{center}
\subfigure[\enskip fit quality.]{
   \includegraphics[height=0.19\textheight]{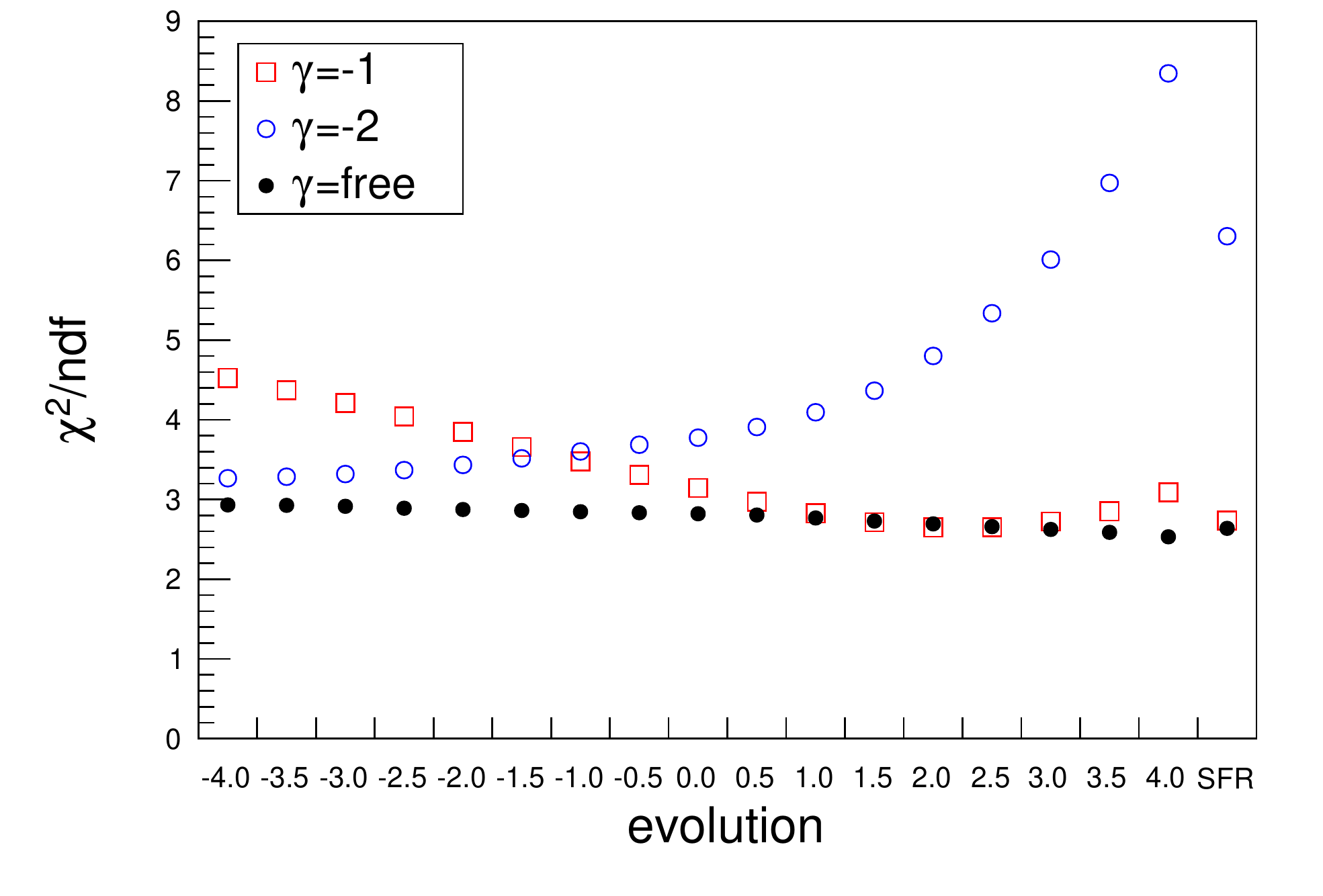}
   \label{fig:gamma0}
}
\subfigure[\enskip peak energy.]{
   \includegraphics[height=0.19\textheight]{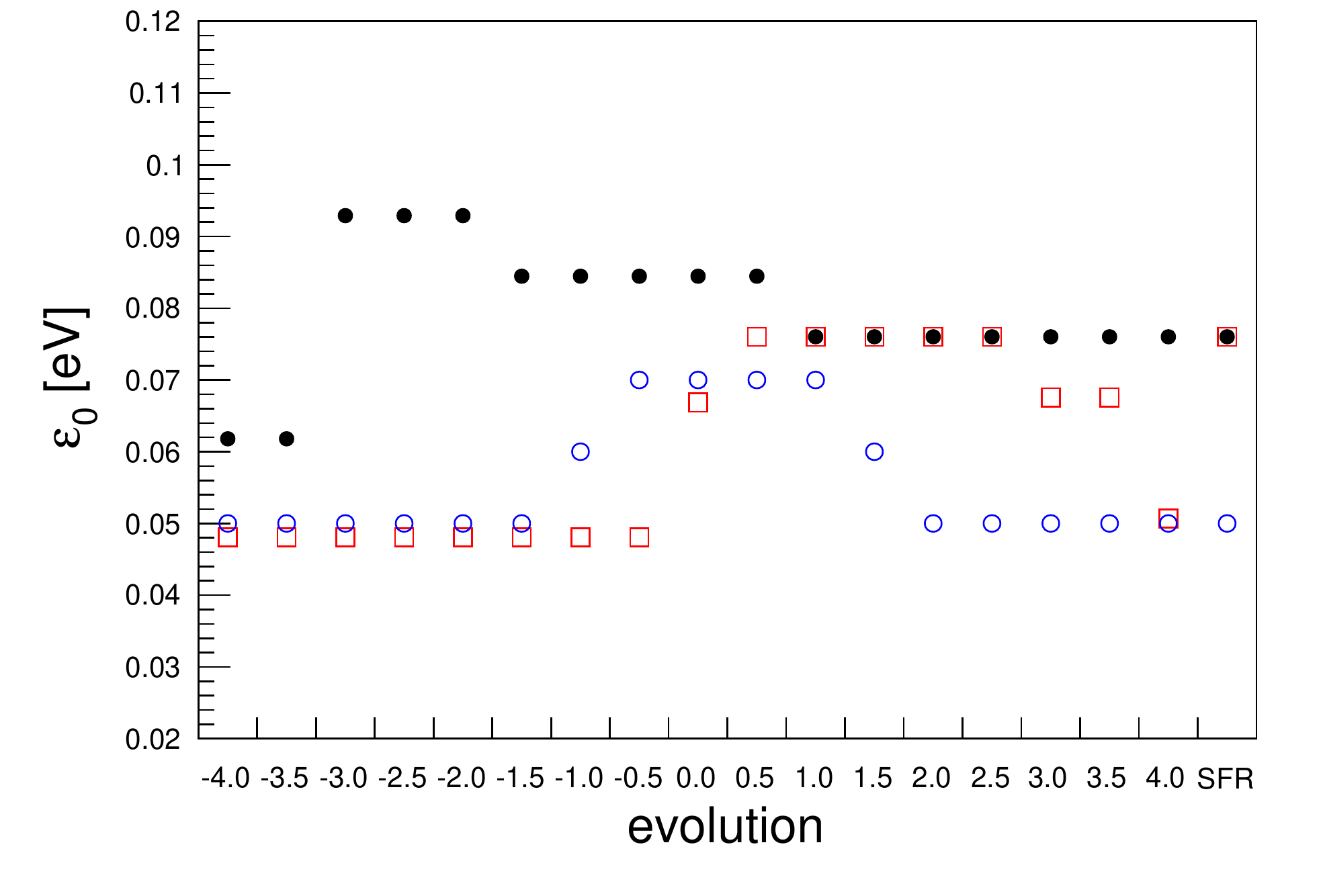}
   \label{fig:gamma1}
}
\subfigure[\enskip spectral index of injected spectrum.]{
   \includegraphics[height=0.19\textheight]{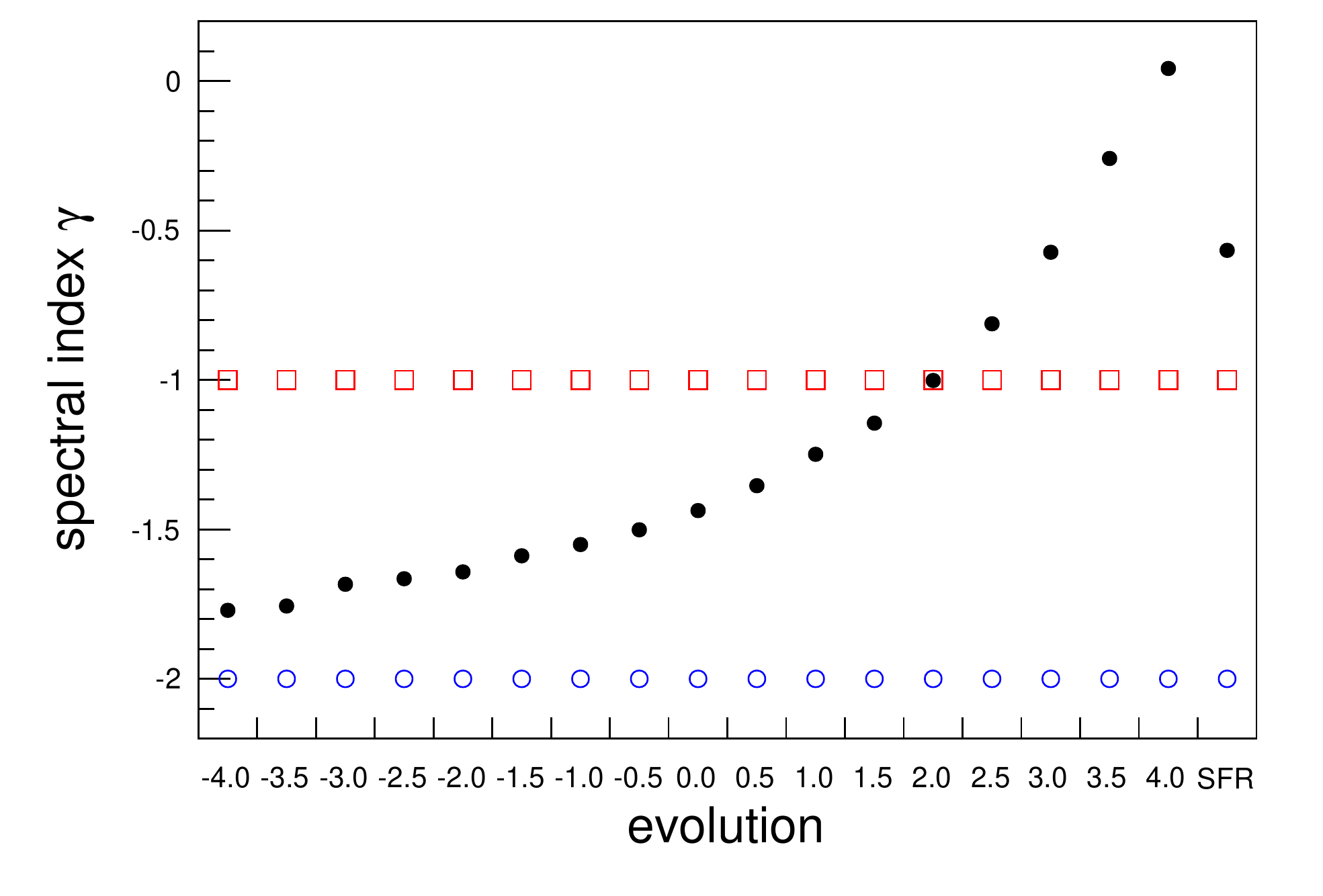}
   \label{fig:gamma2}
}
\subfigure[\enskip ratio of escape and interaction time.]{
   \includegraphics[height=0.19\textheight]{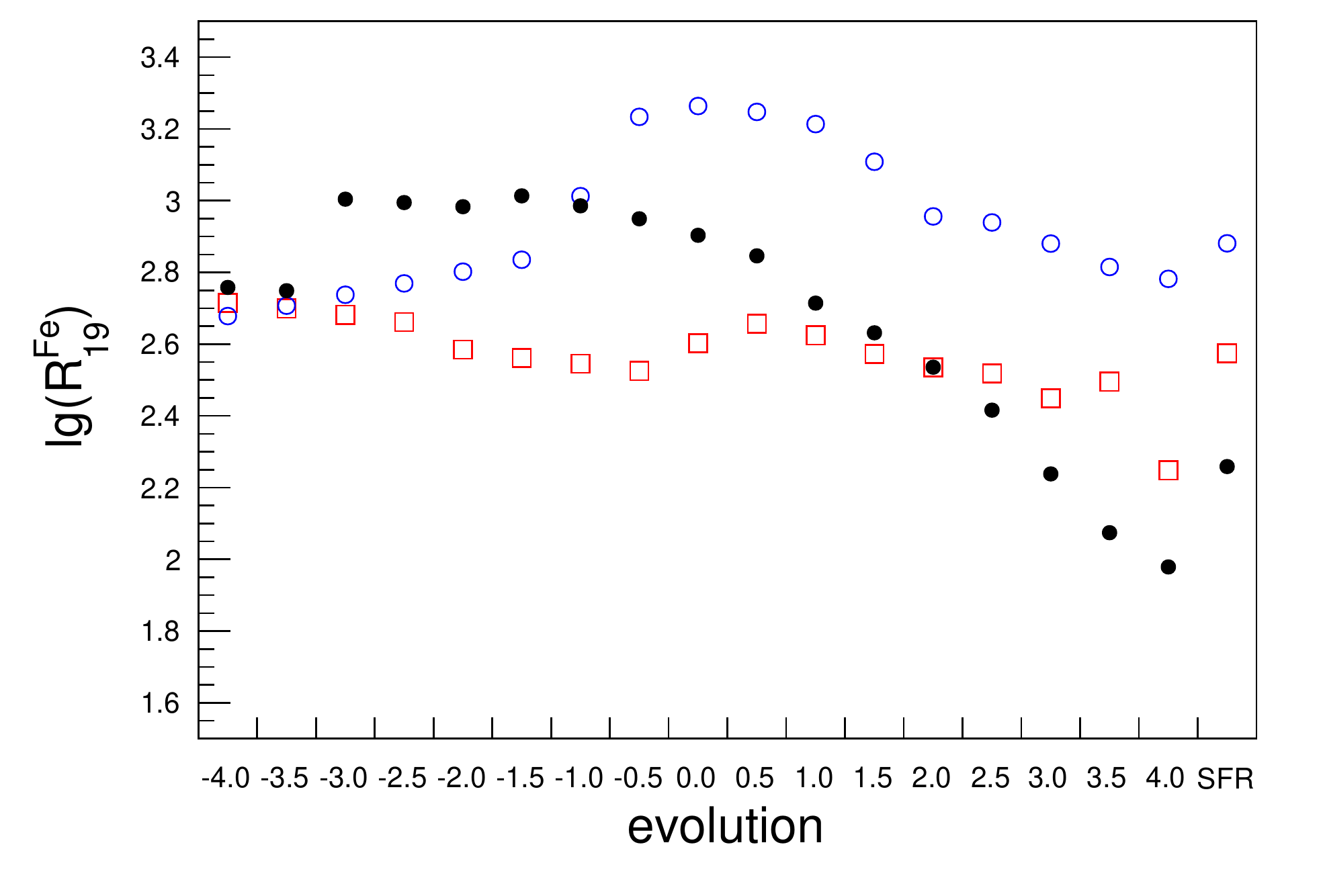}
   \label{fig:gamma3}
}
\subfigure[\enskip power-law index of escape time.]{
   \includegraphics[height=0.19\textheight]{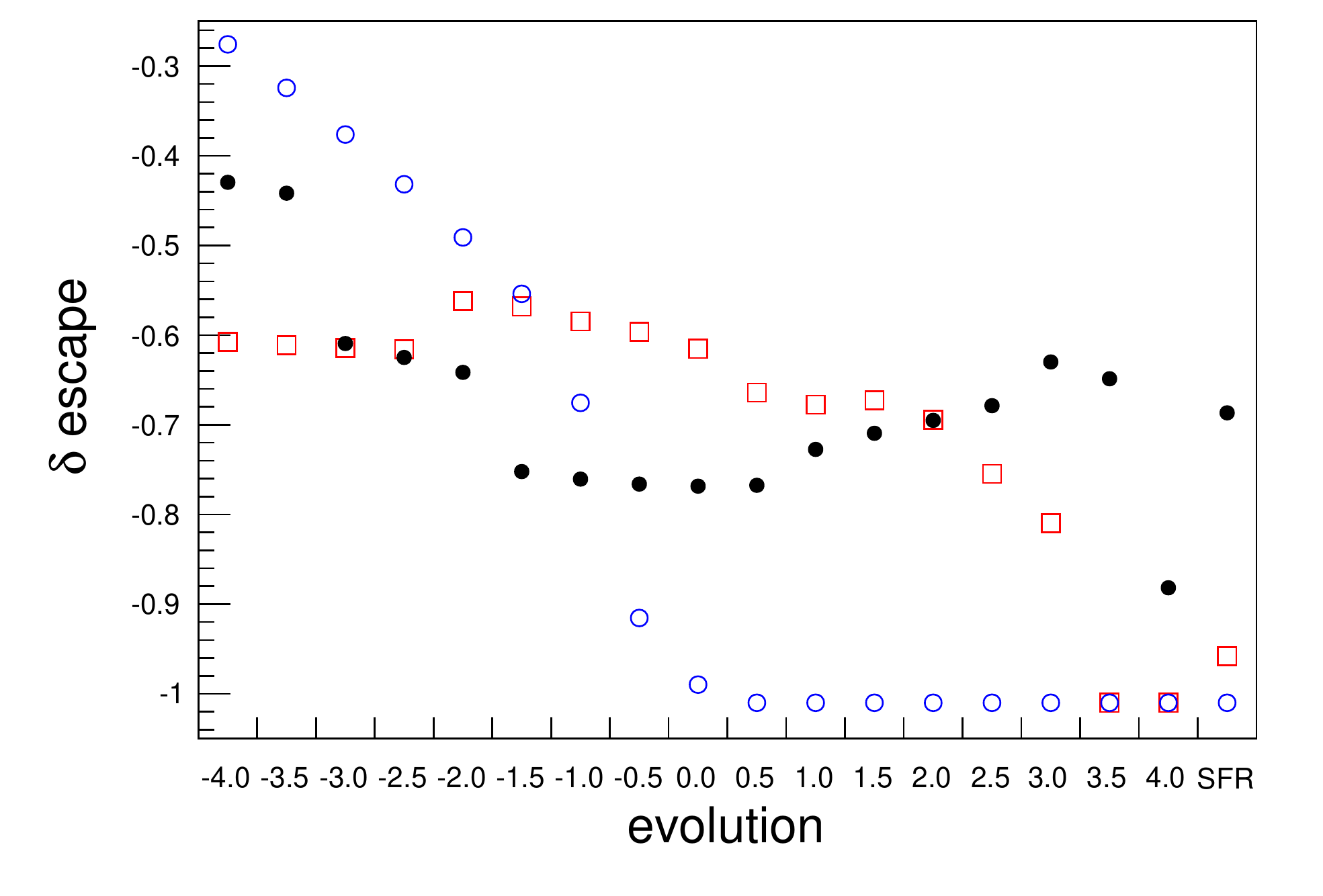}
   \label{fig:gamma4}
}
\subfigure[\enskip maximum energy for $Z=1$.]{
   \includegraphics[height=0.19\textheight]{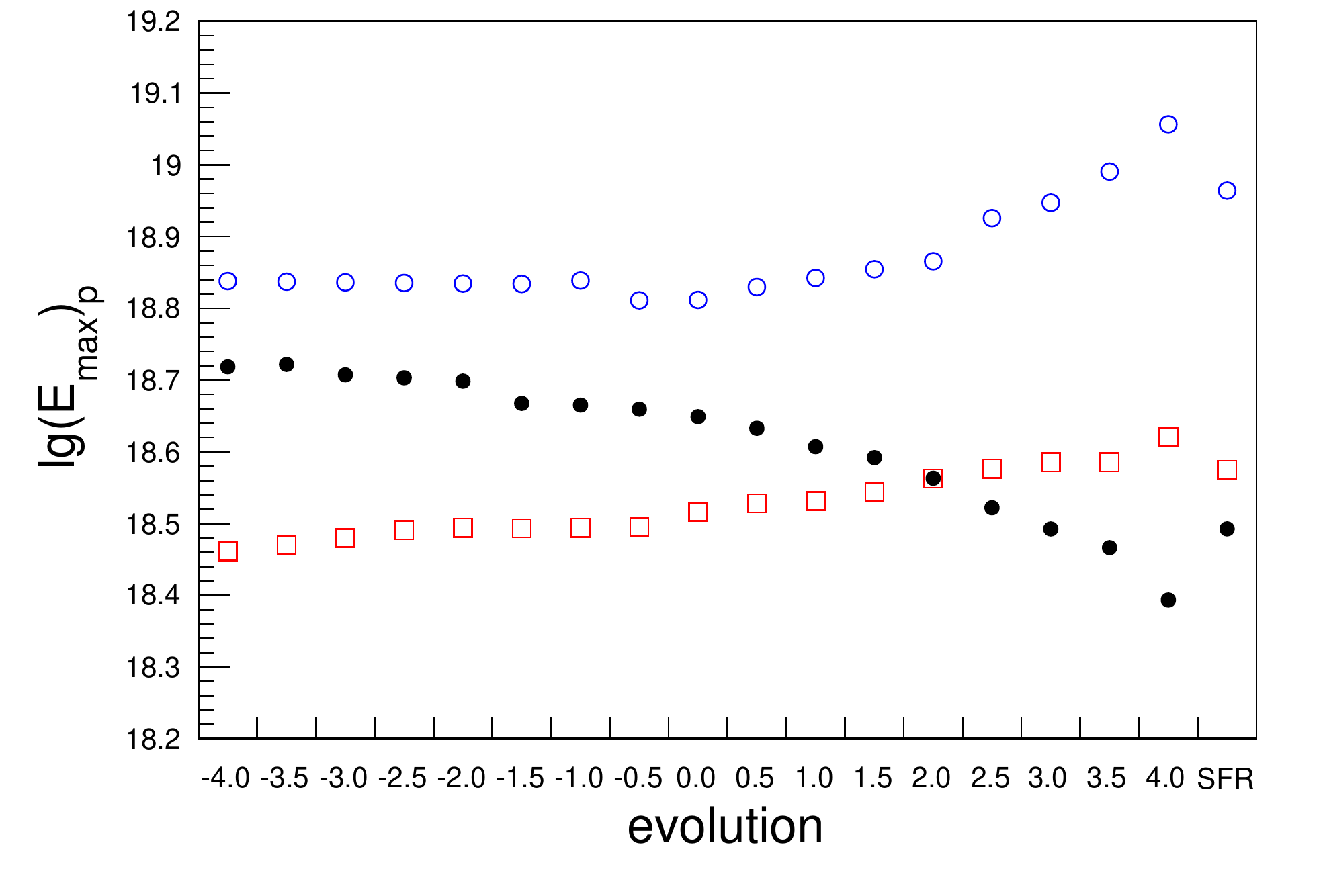}
   \label{fig:gamma5}
}
\subfigure[\enskip injected mass]{
   \includegraphics[height=0.19\textheight]{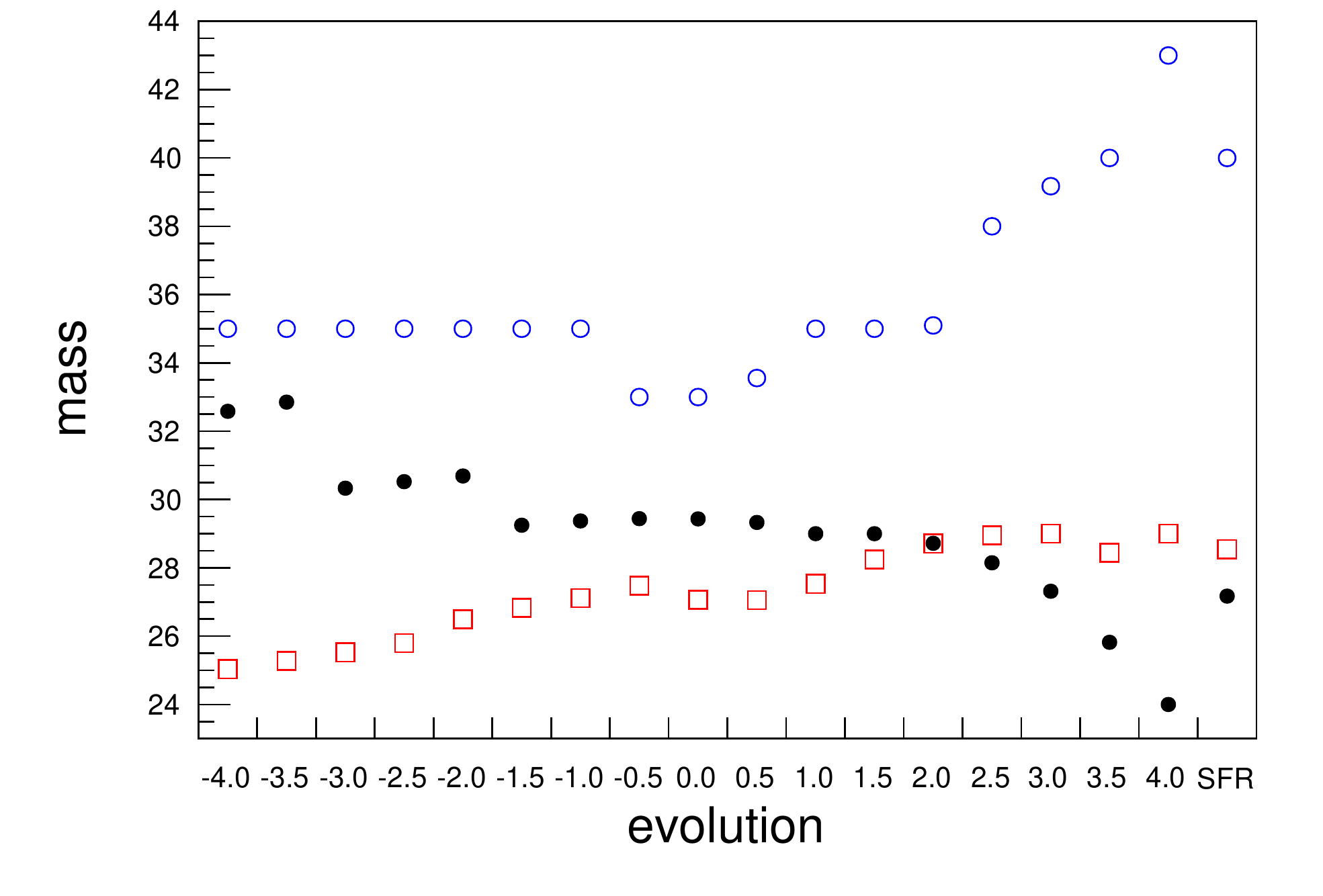}
   \label{fig:gamma7}
}
\subfigure[\enskip UHECR energy injection rate]{
   \includegraphics[height=0.19\textheight]{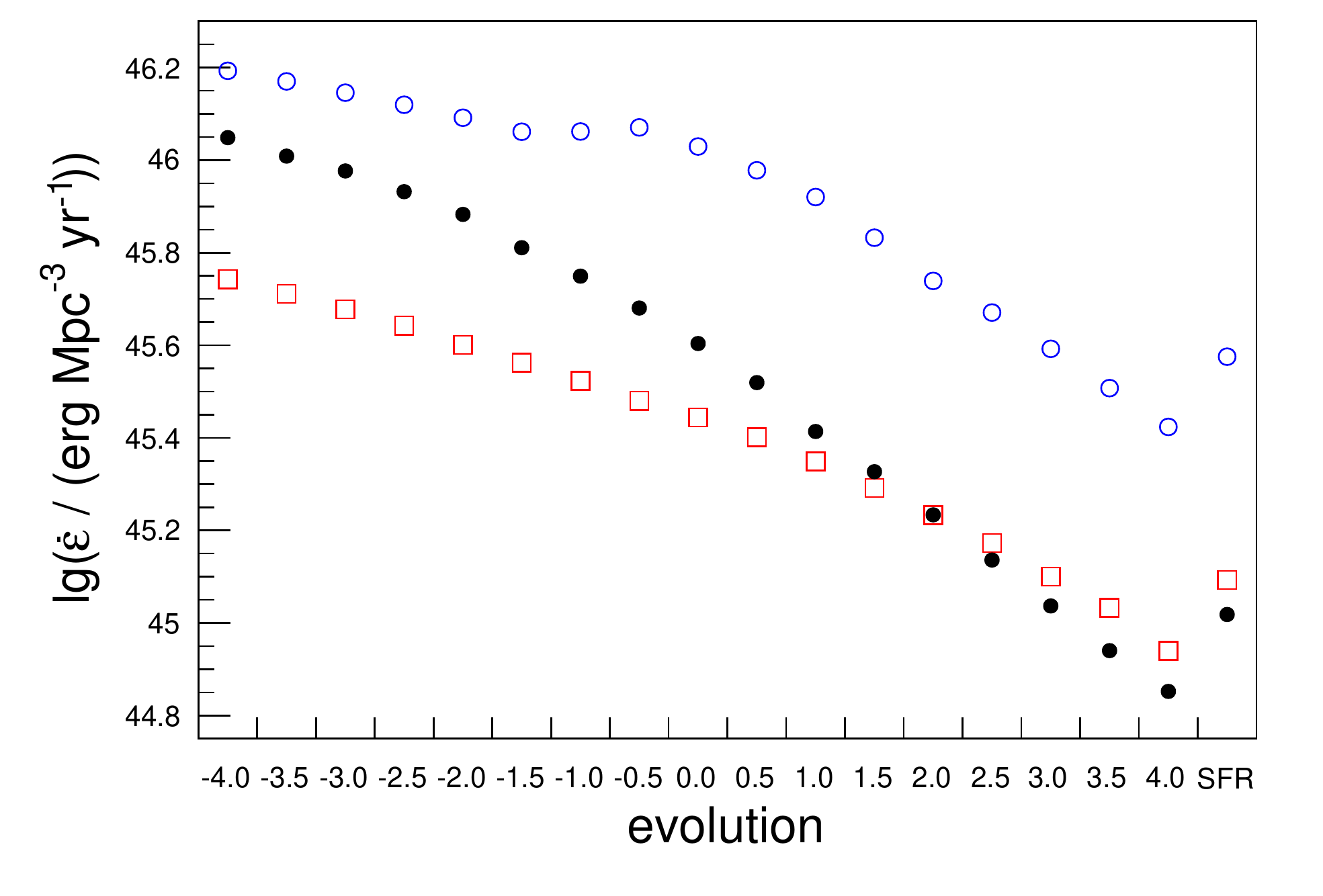}
   \label{fig:gamma9}
}
\end{center}
\caption[fitGamma]{Fit results as a function of source evolution for
  different spectral indices of the injected flux: $\gamma$ fixed to $-1$ (open
  squares), fixed to $-2$ (open circles) and best fit (filled
  circles). On the x-axis the power $m$ of the source evolution is
  shown; the last bin reports the values for the fiducial model (SFR) evolution
  from~\cite{Robertson:2015uda}, Eq.\,(\ref{eq:evolution}).}
\label{fig:gammaFits}
\end{figure*}

To have a concrete fiducial model, we needed to specify how the production of UHECRs varied over cosmological time scales.  This is known as the source evolution, which we took to be in direct proportion to the star-formation-rate -- as would be expected in a source scenario such as young magnetars.  In this section, we consider alternative evolutions of the source luminosity density described by the simple one-parameter functional
form
\begin{equation}
\xi(z) =
\begin{cases}
    (1+z)^m & z < z_0 \\
    (1+z_0)^m \, \exp\left(-(z-z_0)\right) & \text{otherwise}
\end{cases}
\end{equation}
with $z_0=2$ and $m$ ranging from $-4$ to $+4$.
$m=0$ yields a uniform
source luminosity distribution, $m=+4$ corresponds to a strong evolution similar
to the one of active galactic nuclei, and negative values result in sources that
are most abundant or most luminous within the low-redshift universe as
suggested in~\cite{Taylor:2015rla}. The resulting fit parameters are
displayed in Fig.\,\ref{fig:gammaFits} for three choices of the spectral
index $\gamma$ of the injected flux: $-1$, as in the fiducial model,
$-2$ as expected for stochastic shock acceleration and for letting
$\gamma$ float freely in the fit. As can be seen in Fig.\,\ref{fig:gamma0},
 $\gamma = -2$  gives a poor description of the data for $m\gtrsim 0$, but is a
viable choice for closeby sources, in accordance to the findings of~\cite{Taylor:2015rla}.
For positive values of $m$, a fixed value of $\gamma=-1$ gives a similar fit quality
as the freely floating $\gamma$, but the latter converges to values larger than $-1$
for source evolutions with $m>2$ (cf.\ Fig.\,\ref{fig:gamma2}).

For the ``traditional'' source evolutions with $m \geq 0$ and the fit
with $\gamma=-1$ we find that most of the parameters exhibit only a
minor variation with $m$, with the exception of the power-law index of
the escape time $\delta$ (Fig.\,\ref{fig:gamma4}) and the power density
$\bolddot{\epsilon}_{17.5}$ (Fig.\,\ref{fig:gamma4}).

We conclude that our model for the ankle does not critically depend on
the choice of the source evolution, but that for a given choice of $m$
we can constrain the allowed values of $\gamma$, $\delta$ and
$\bolddot{\epsilon}_{17.5}$.

\subsubsection{Photon Spectrum}
\label{sec:modelVarPhotonSpec}
\begin{figure*}[p!]
\begin{center}
\subfigure[\enskip fit quality.]{
   \includegraphics[height=0.19\textheight]{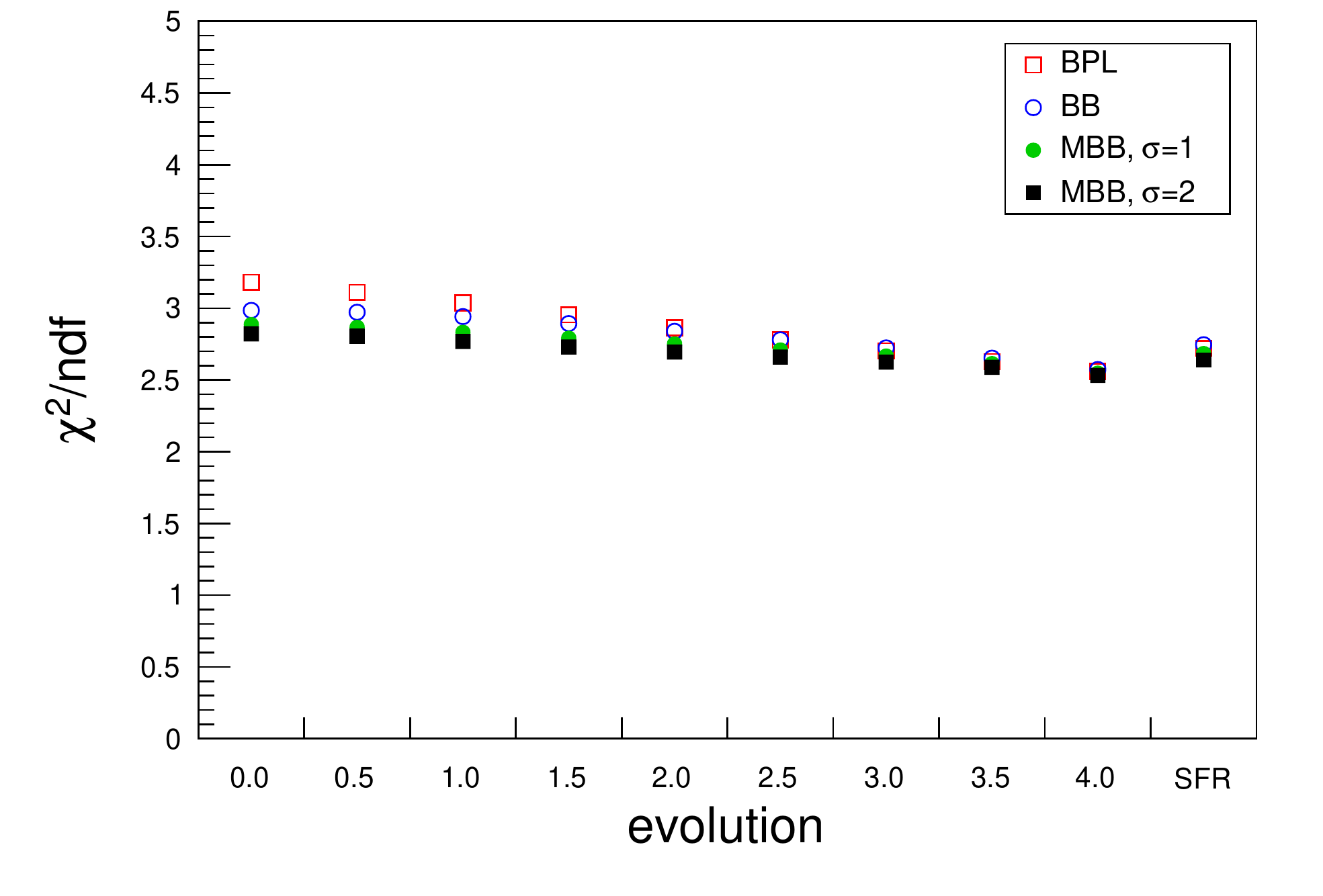}
   \label{fig:photon0}
}
\subfigure[\enskip peak energy.]{
   \includegraphics[height=0.19\textheight]{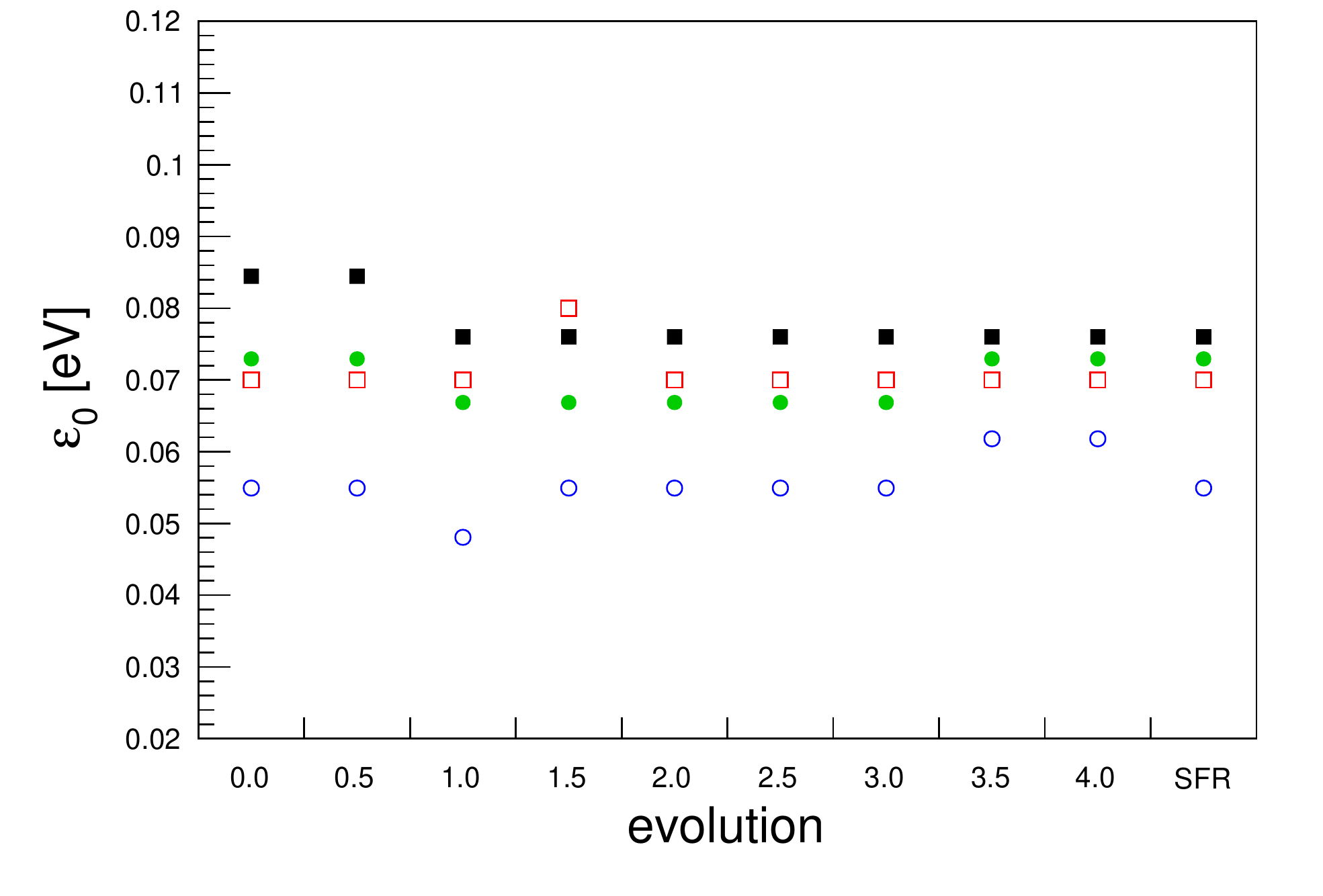}
   \label{fig:photon1}
}
\subfigure[\enskip spectral index of injected spectrum.]{
   \includegraphics[height=0.19\textheight]{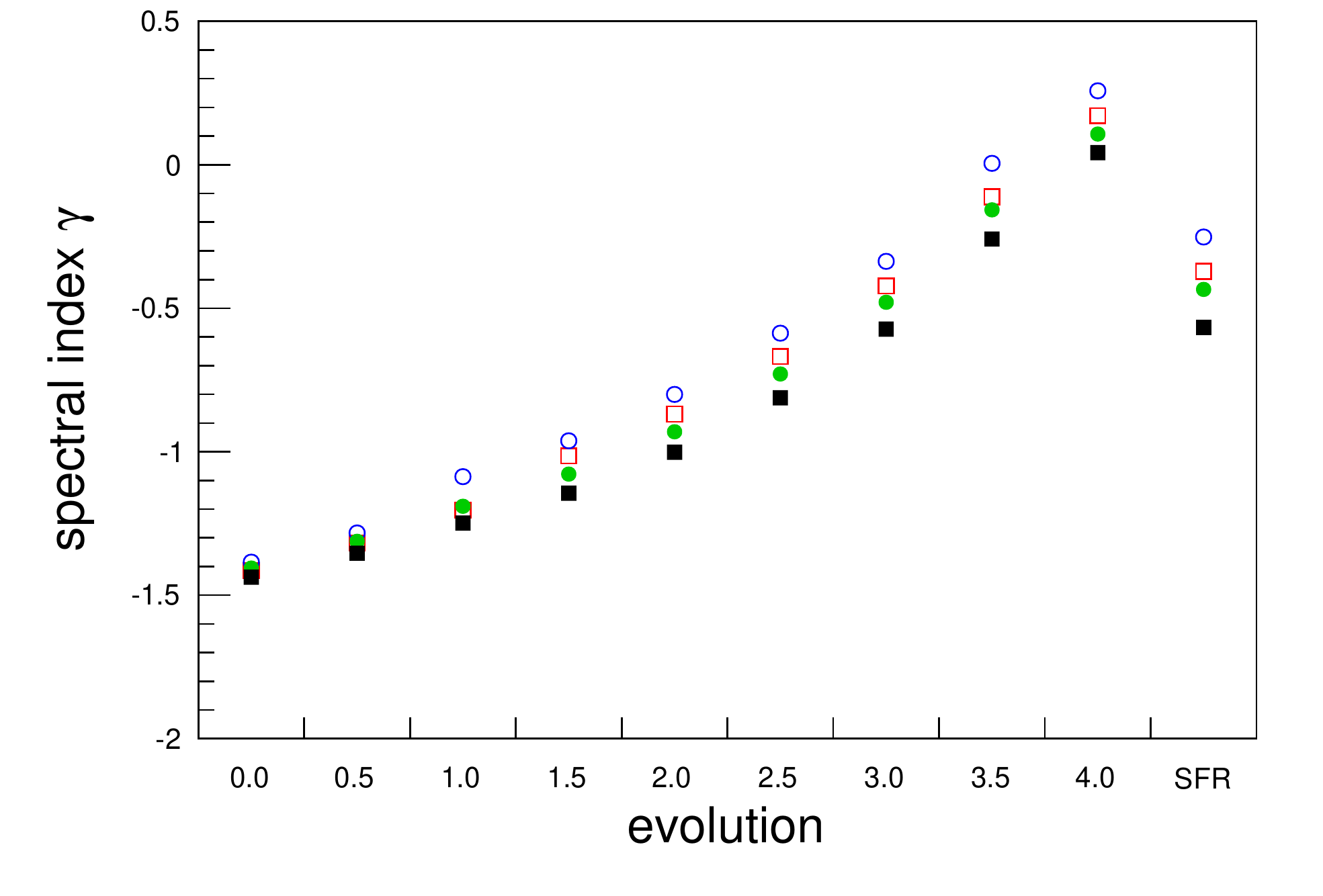}
   \label{fig:photon2}
}
\subfigure[\enskip ratio of escape and interaction time.]{
   \includegraphics[height=0.19\textheight]{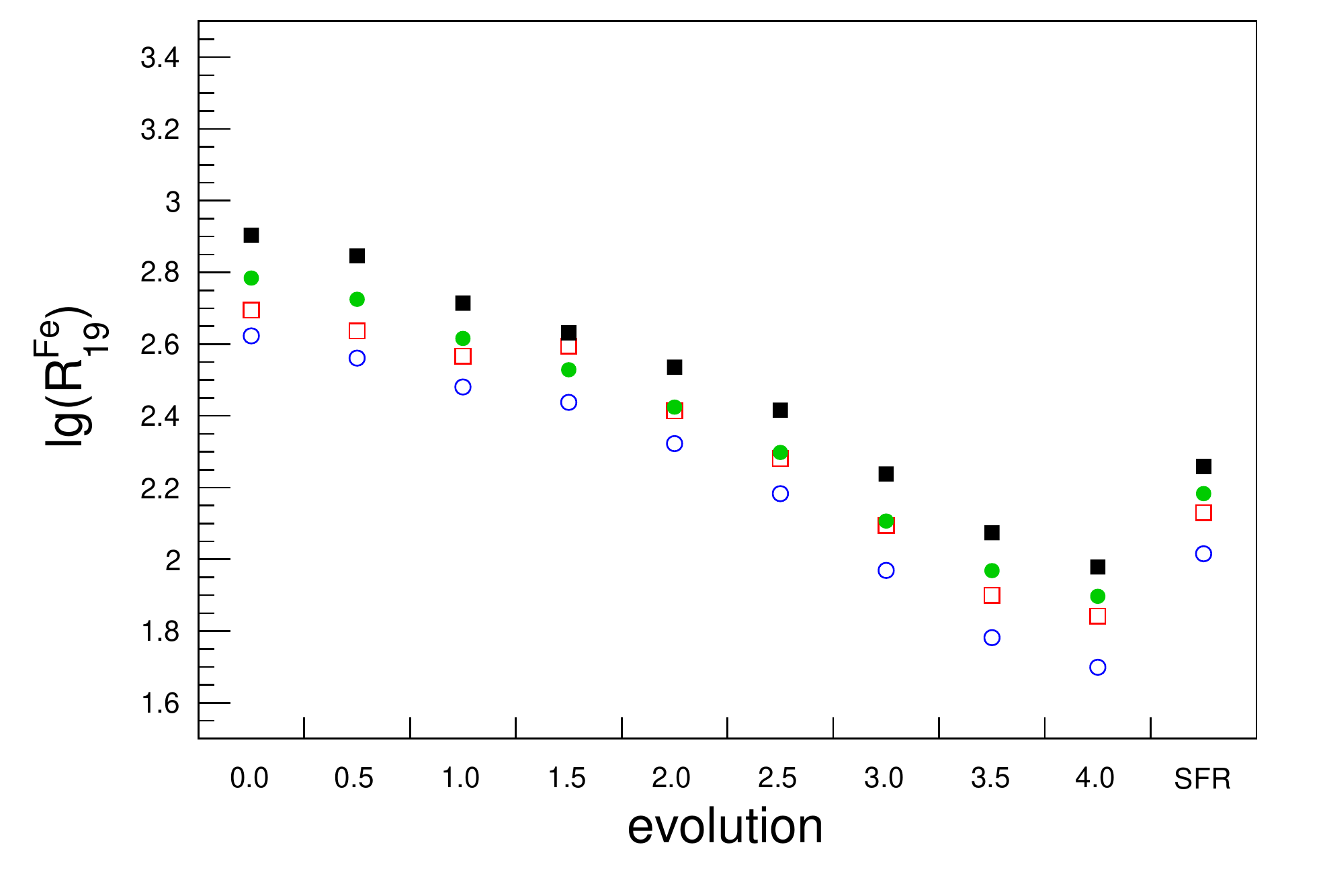}
   \label{fig:photon3}
}
\subfigure[\enskip power law index of escape time.]{
   \includegraphics[height=0.19\textheight]{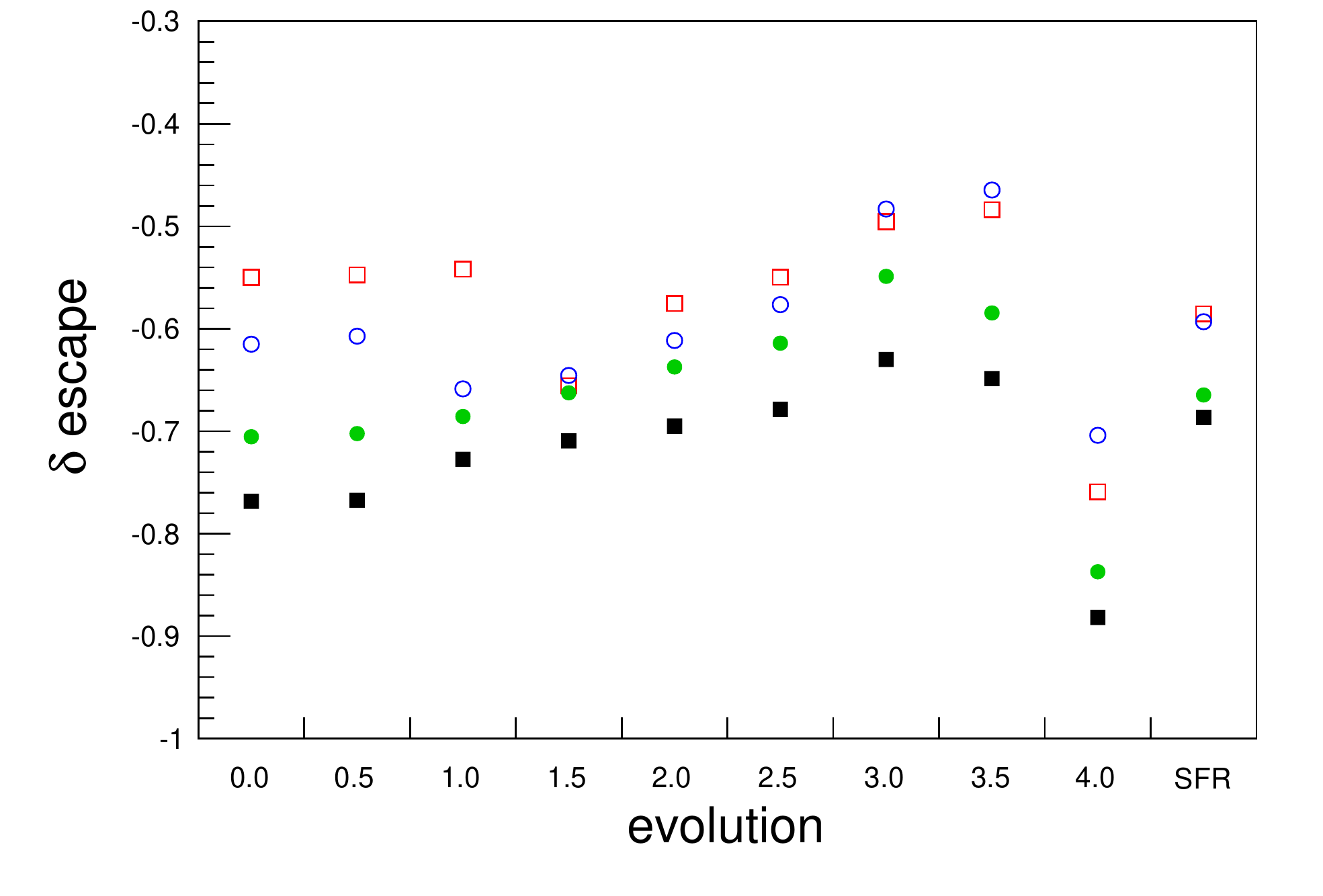}
   \label{fig:photon4}
}
\subfigure[\enskip maximum energy for $Z=1$.]{
   \includegraphics[height=0.19\textheight]{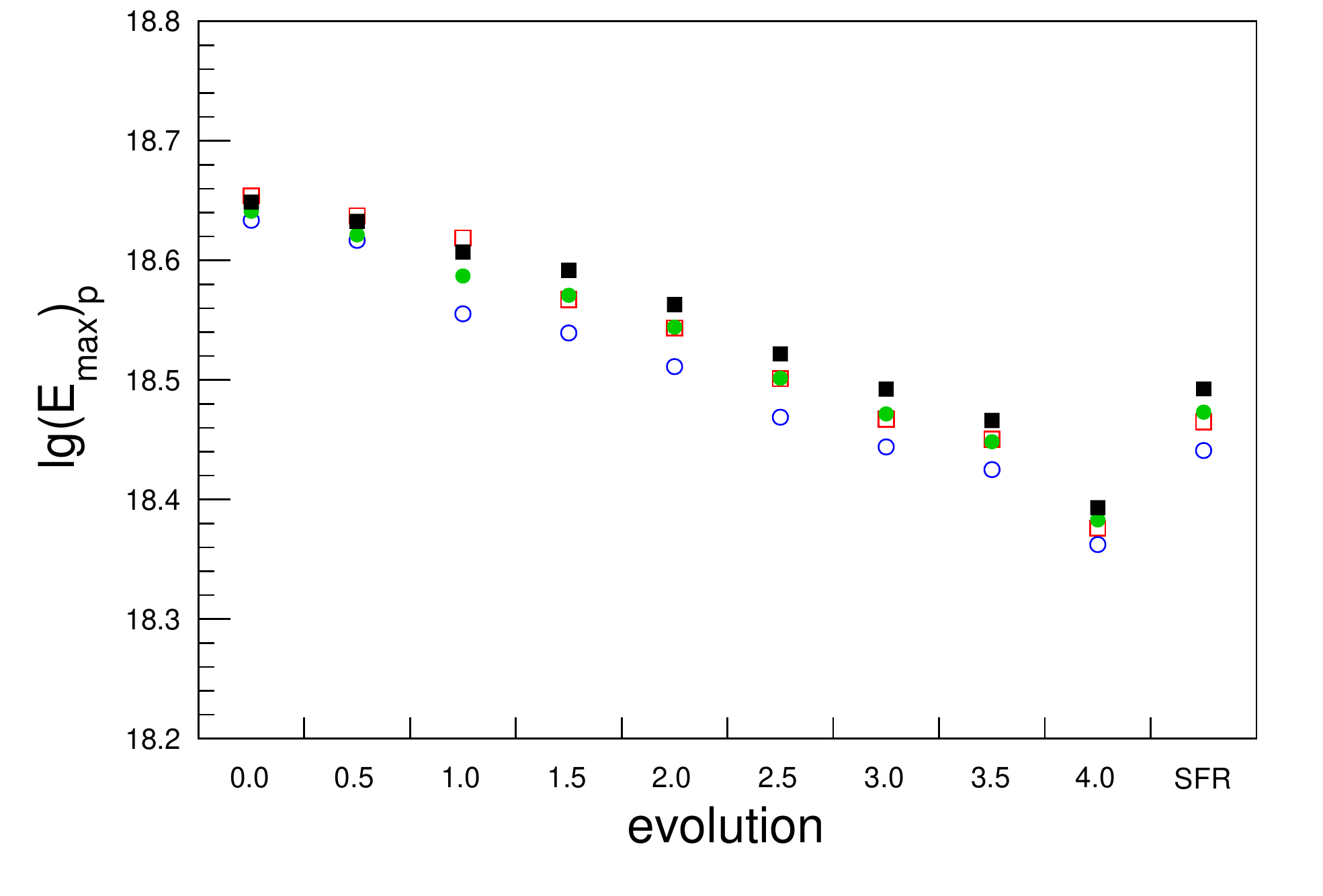}
   \label{fig:photon5}
}
\subfigure[\enskip injected mass]{
   \includegraphics[height=0.19\textheight]{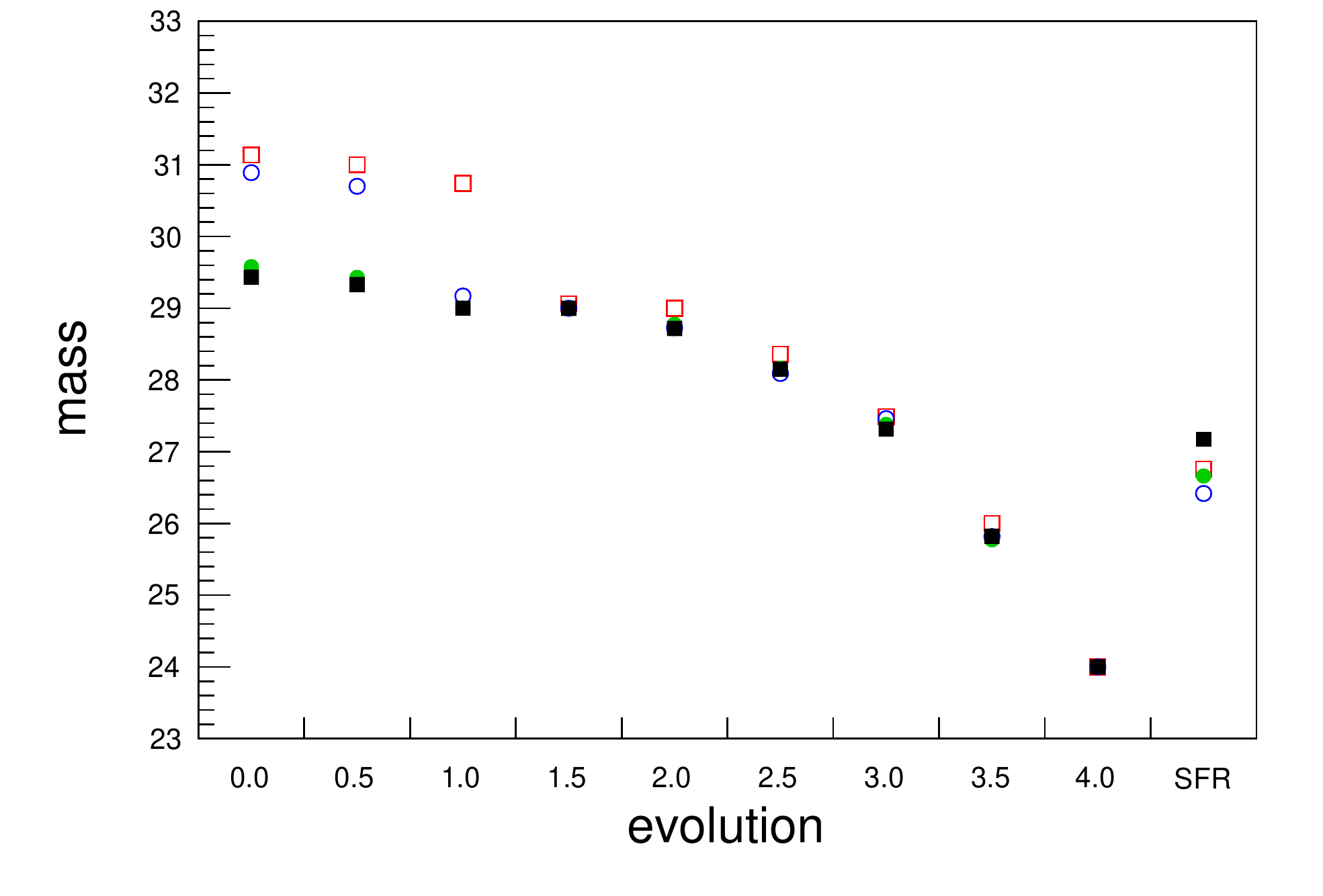}
   \label{fig:photon7}
}
\subfigure[\enskip UHECR energy injection rate]{
   \includegraphics[height=0.19\textheight]{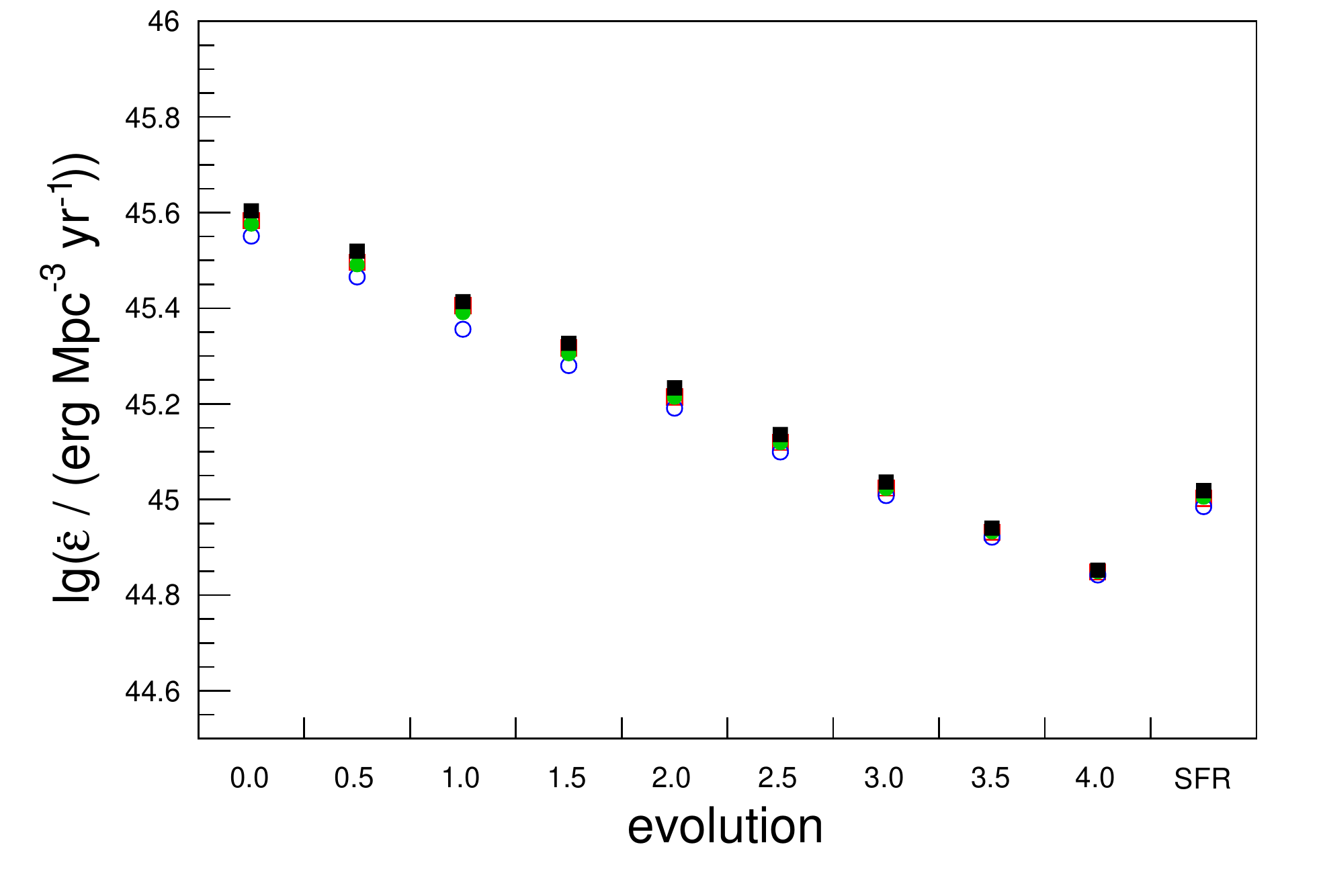}
   \label{fig:photon9}
}
\end{center}
\caption[fitPhoton]{Fit results as a function of source evolution for
  different photon spectra: Broken power law (BPL, open squares),
  black body spectrum (BB, open circles), modified black body spectrum
  (MBB) with $\sigma=1$ (filled circles) and $\sigma=2$ (filled
  squares). On the x-axis the power $m$ of the source evolution is
  shown and in the last bin the fit values for the fiducial evolution
  from~\cite{Robertson:2015uda}, Eq.\,(\ref{eq:evolution}), is shown.
}
\label{fig:photonFits}
\end{figure*}

We repeated the model fits using alternative energy distributions of
the photon density instead of the broken power law used in the
fiducial model: a black body spectrum and two modified black body
spectra. All four spectra are normalized to the same integral photon
density and depend only on one parameter, the peak energy
$\varepsilon_0$ (see Appendix~\ref{app:appendixA}).  The resulting
fit results are shown in Fig.\,\ref{fig:photonFits} for a freely
floating spectral index $\gamma$ and for source evolutions with $m
\geq 0$. As can be seen, all four photon spectra describe the data
equally well (Fig.\,\ref{fig:photon0}). The best-fit values of the
free model parameters are very similar and in particular the obtained
peak values are within $\pm 20$~\meV.
We conclude that as long as the photon spectrum is ``peaky'', the
particular details of its shape do not influence the parameters of
our model.

\begin{figure*}[thb]
\centering
 \includegraphics[width=0.48\linewidth]{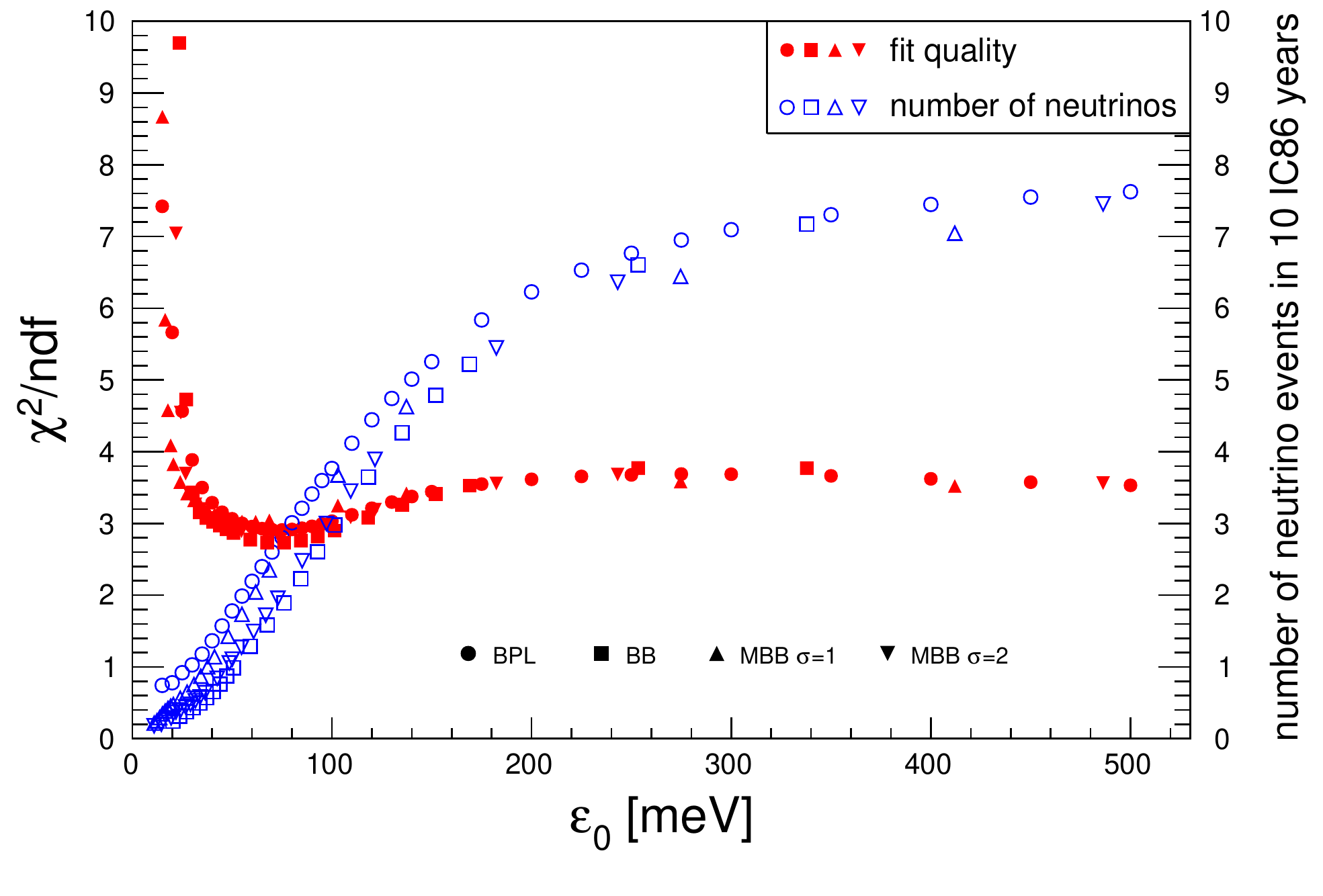}
 \includegraphics[clip, viewport = 9 380 490 739, width=0.492\linewidth]{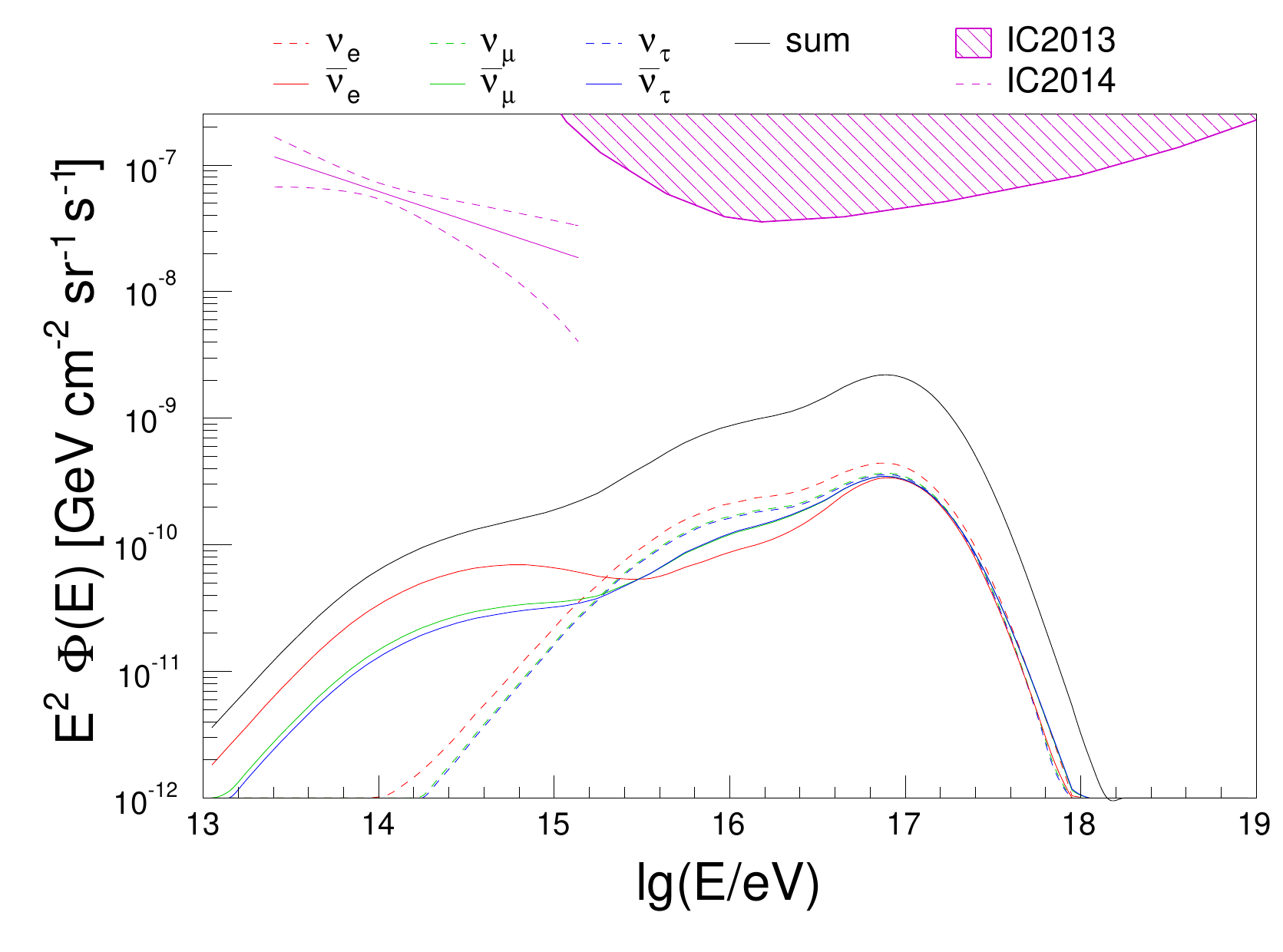}
\caption[chi2andnu]{Left: Fit quality of the fiducial model (closed symbols) and number of
  neutrinos (open symbols) as a function of peak energy
  $\varepsilon_0$ of the photon spectrum in the source environment.
  Four types of photon spectra are shown: Broken power law (BPL),
  black body spectrum (BB) and two modified black body spectra (MBB).
  The minimum $\chi^2$ of BPL corresponds to the result shown in
  Fig.\,\ref{fig:fitSys}. Right: Lower limit on the neutrino flux
  obtained for a modified black body spectrum with $\sigma=2$, and
  $\varepsilon_0 = 34$~\meV ($T = 100$~K).  The lines and hatched area
 at the top of the figure are the measured neutrino flux and upper limit
 from IceCube~\cite{Aartsen:2014muf, Aartsen:2013dsm}}
\label{fig:chi2andnu}
\end{figure*}

The sensitivity of the fit to the peak energy is shown in the left
panel of Fig.\,\ref{fig:chi2andnu}. As can be seen, the $\chi^{2}$
deteriorates very quickly at low values of $\varepsilon_0$, but it is almost flat above
the minimum. This feature can be easily understood recalling $\varepsilon_0$ in the
``L-curve'' approximation introduced in Sec.\,\ref{sec:ankle}: The
smaller $\varepsilon_0$, the larger is the energy of inflection of the interaction length,
$E_b$. For too-small values of $\varepsilon_0$, the interaction and escape times are parallel over
the full energy range and thus no high-pass filter is created. On the
other hand, once $E_b$ is small enough, a further decrease changes
only the flux at low energy, where the escaping spectrum is
dominated by low-mass nuclei from spallation (see
e.g.\ Fig.\,\ref{fig:tau&escspec}) which can be compensated by adjusting other parameters such as $R_{19}^{\rm Fe}$.

To first order, our model can therefore only give a lower limit on the
peak energy of the photon flux in the source environment. However, future limits or observations of
neutrinos in the 10-100~\PeV range will help to constrain this important
source property, because the number of predicted neutrinos strongly
depends on $\varepsilon_0$, as shown in the left panel of
Fig.\,\ref{fig:chi2andnu} by the superimposed open symbols. A larger
peak energy of the ambient photon environment increases neutrino production at the source in two
ways. Firstly, shifting $E_b$ to lower energies (and compensating as necessary by adjustment of
$R_{19}$) moves the interaction times of protons closer to the
escape time and correspondingly additional neutrinos are produced via
photo-pion production of protons (compare e.g. the red curves at
around \energy{18} in the upper panel of Fig.\,\ref{fig:tau&escspec}
($\varepsilon_0 = 110$~\meV) to the ones in Fig.\,\ref{fig:tauSys}
($\varepsilon_0 = 70$~\meV)). Secondly, increasing $\varepsilon_0$
moves the minimum of the interaction time for photo-pion production of
nuclei to lower energies. Since the neutrinos from photo-pion
production carry a larger fraction of the nucleon energy than the
neutrinos from neutron decay after photo-dissociation, this increases
the neutrino flux as well.

It is tempting to give a quantitative interpretation of the
$\chi^2$-curve of Fig.\,\ref{fig:chi2andnu} in terms of a lower limit
on $\varepsilon_0$ and the number of neutrinos. However, the minimum
of $\chi^2$ is far away from $\chi^2/{N_{\rm df}}=1$ which -- assuming
this model is correct -- is indicative of experimental systematics or an under-estimation of the
experimental uncertainties or of deficiencies in the modeling of
hadronic interactions in the atmosphere needed to interpret the data
in terms of mass composition (see above).  In the absence of a
concrete explanation we follow the
PDG~\cite{Rosenfeld:1975fy,Agashe:2014kda} and rescale the
uncertainties by a common factor $S = (\chi^2_{\rm min} /{N_{\rm
    df}})^\frac{1}{2}$ to bring the rescaled  $\chi^2/{N_{\rm df}}$ to 1.  This rescales the $\chi^2$ value of any given model so that the number of standard-deviations it is from the minimum is given by $N_\sigma^\prime = S^{-1} \,
\sqrt{\chi_{\rm model}^2 -\chi^2_{\rm min}}$.  This yields an approximate lower
limit on $\varepsilon_0$ at $N_\sigma^\prime = 3$ of  $\varepsilon_0 > 34$~\meV and
$N_\nu(10\times \rm IC86) > 0.4$ assuming the validity of the fixed
fiducial parameters given in Table~\ref{table:fiducial}. The
corresponding lower temperature limits are 180~K, 125~K and 100~K for
the black body spectra with $\sigma=0$, 1 and 2 respectively.  The
lower limit on the neutrino spectrum is shown in the right panel of
Fig.\,\ref{fig:chi2andnu}.

\subsubsection{Hadronic Interactions in the Source Environment}
\label{sec:hadronicAtSource}
In addition to interactions with the background photon field, nucleons
and nuclei can also scatter off hadrons in the source environment. In
this paper we assume that the density of hadronic matter in the source environment is low enough that such hadronic interactions can be neglected.  For any concrete
astrophysical realization of our scenario, one must check and if
necessary include hadronic interactions in the source environment.
Production of $\pi^\pm$'s and $\pi^0$'s in
hadronic collisions could significantly increase the fluxes of
neutrinos and photons emitted in the EeV energy range. Fast-spinning
newborn neutron stars provide a particular
example~\cite{Fang:2013vla}. Precise estimates of the impact of
hadronic collisions on the predictions of our model will be presented
in a separate publication.  The results presented here are valid for
all astrophysical systems in which the interactions are dominated by
photo-nuclear processes.

\section{Conclusions}
\label{sec:conclusions}

In this paper we have proposed a new explanation for the ankle in the cosmic ray spectrum, and for the evolution with energy of the composition of extragalactic cosmic rays: from light below the ankle to increasingly heavy above.
When nuclei are trapped in the turbulent magnetic field
of the source environment, their escape time can decrease faster with
increasing energy than does their interaction time.  Under these
conditions, only the highest energy particles can
escape the source environment unscathed, and the source
environment acts as a high-pass filter on UHECRs.
Nuclei below the crossover energy such that $\tau_{\rm esc} > \tau_{\rm int}$ interact with photons
in the environment around the source, with ejection of nucleons or alpha
particles and consequent production of a steep spectrum of secondary nucleons.  The
superposition of this steeply falling nucleon spectrum with the harder
spectrum of the surviving nuclear fragments creates an ankle-like
feature in the total source emission spectrum.    Above the ankle, the spectrum emerging from the source environment exhibits a progressive transition to heavier nuclei, as the escape of non-interacting nuclei becomes efficient.  Abundant
production of $\bar{\nu}_e$'s is a signature of this mechanism.

We illustrated the high quality of the fit which can be obtained to
the Auger data, with a fiducial model in which nuclei are accelerated
up to a maximum rigidity found to be $\approx 10^{18.5}$ V, with
spectrum $\propto E^{-1}$, and are then subject to
photo-disintegration in the vicinity of the accelerator before
escaping for their journey to Earth.  We showed that the details of
the photon spectrum around the accelerator are unimportant, except for
its peak energy.  The other important characteristic of the
environment is the photon density relative to the magnetic
diffusivity, which we characterized in a very simplistic way (through
a single parameter) in this initial study.  We studied the sensitivity
of the mechanism to the energy-scale uncertainty and
hadronic-interaction-modeling uncertainty, which affects the
composition inferred from the atmospheric shower observations, and
also used the TA spectrum instead of the Auger spectrum.  The
conclusion of these studies is that a good quality fit can be obtained
in most cases, but details of the fit parameters such as the
composition and maximum energy characterizing the accelerator change.
A corollary is that until these systematic uncertainties in the
observations and their interpretation are reduced, such details of the
accelerator cannot be reliably inferred from the data.  The fiducial
model parameters needed in the fits are such that the scenario can be
reasonably achieved in at least one type of proposed astrophysical
source, as will be discussed in a future publication.

Our mechanism has two predictions beyond fitting the shape of the
spectrum and composition evolution, which are independent of many
environmental variables and can be used to test the validity of this
scenario for production of the ankle.  i) The spectral cutoff of
spallated nucleons emerging from the source environment is
$\frac{1}{2} R_{\rm max}$, where $R_{\rm max}$ is the rigidity cutoff
of the accelerator, because $E_{\rm max, \, spal. nuc.} = E_{{\rm
    max}, A}/ A$ while $E_{{\rm max}, A} = Z \, R_{\rm max}$, and
finally $Z/A = \frac{1}{2}$, largely independent of composition.  This
relation holds prior to the extragalactic propagation from the source,
thus giving complementary information on the accelerator to that
obtained from the spectrum and composition above the ankle alone.  ii)
There is a one-to-one relation between the spectrum of spallated
nucleons and the anti-electron-neutrinos produced by beta decay of
neutrons, unless the spallated nucleons lose energy by interacting
with hadronic material in the source environment.  Independent of
other properties of the environment or the source evolution,
$\bar{\nu}_e$'s will have an identical spectral shape, shifted down by
a factor $\sim 1/1000$ from the kinematics of $n\rightarrow p \, e^{-}
\bar{\nu}_{e}$ and reduced by a factor-2 in normalization because only
half the nucleons are neutrons.  This follows because propagation
energy losses are small for nucleons of such low energy, and redshift
impacts both nucleons and neutrinos identically.  Thus, detailed
comparison of the $\bar{\nu}_e$ and spallated nucleon spectra will
reveal if hadronic interactions in the source environment are
important, which would imply a correlated production of photo-pion
produced neutrinos.

\section*{Note Added}
After this work was presented at the IceCube Particle Astrophysics
Symposium a paper appeared on the arXiv exploring another mechanism
for producing the ankle, arising in the context of gamma-ray
bursts~\cite{Globus:2015xga}.\\

\section*{Acknowledgments}
We would like to acknowledge many useful discussions with our
colleagues of the Pierre Auger Collaboration. Furthermore we thank
David Walz for his support regarding questions about \CRP and Benoit
Marchand for his help running the calculations on the BuTinah high
performance computing cluster of NYU Abu Dhabi. MU acknowledges the
financial support from the EU-funded Marie Curie Outgoing Fellowship,
Grant PIOF-GA-2013-624803. The research of GRF is supported in part by
the U.S. National Science Foundation (NSF), Grant PHY-1212538 and the
James Simons Foundation; she thanks KIPAC/SLAC for their
hospitality. The research of LAA is supported by NSF (Grant CAREER
PHY-1053663) and NASA (Grant NNX13AH52G); he thanks the Center for
Cosmology and Particle Physics at New York University for its
hospitality.
\vfill

\close@column@grid
  \clearpage
\twocolumngrid


\onecolumngrid
\appendix

\section{Photon spectra}
\label{app:appendixA}

\begin{figure*}[!t]
 \includegraphics[clip, viewport=0 -10 510 398,height=0.36\linewidth]{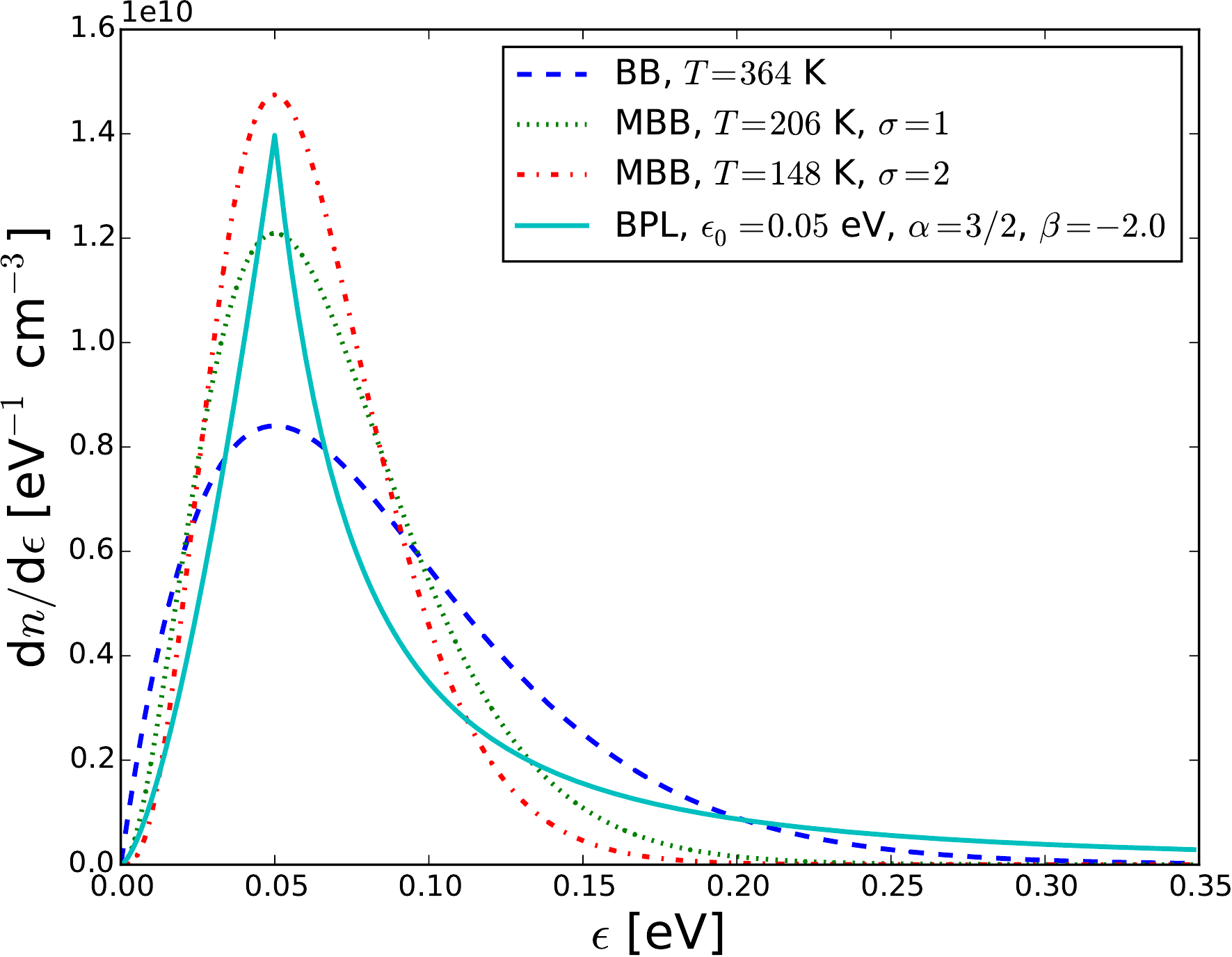}
 \includegraphics[clip, viewport=0 21 550 385, height=0.36\linewidth]{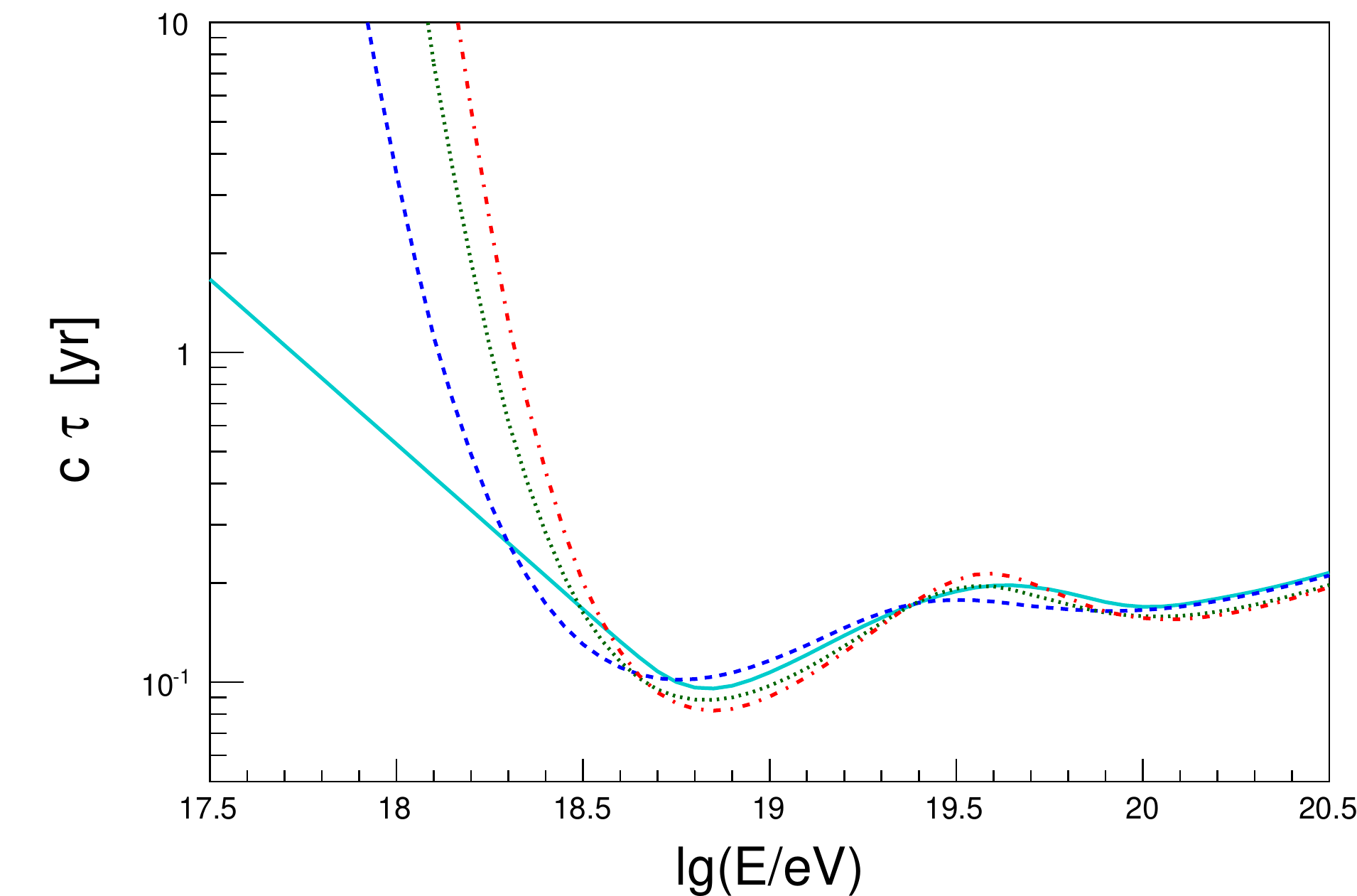}
\caption{Left: Comparison of photon spectra. BPL: Broken power law (solid), BB: black body spectrum (dashed), MBB: modified black body spectrum (dotted and dash-dotted). The curves are normalized to match the integral of the black body spectrum and the temperatures are chosen to match the peak energy of the broken power law. Right: Interaction times corresponding to the four photon spectra.}
\label{app:photon-spectra}
\end{figure*}

In this paper we explore the propagation effects of the four types of
photon spectra shown in Fig.\,\ref{app:photon-spectra}. The first
consists of broken power-law (see e.g.~\cite{Szabo:1994qx}) as a
simplified representative of non-thermal emission given by
\begin{equation}
      n(\varepsilon) = n_0^{\rm BPL}
        \begin{cases}
           (\varepsilon/\varepsilon_0)^\alpha & \varepsilon < \varepsilon_0 \\
           (\varepsilon/\varepsilon_0)^\beta & \text{otherwise} \, .
        \end{cases}
\label{app:eq:photonfield}
\end{equation}
where $\varepsilon$ is the photon energy and the maximum of the number density is at
an energy of $\varepsilon_0$.

We also consider modified black-body spectra using the functional
form
\begin{equation}
      n(\varepsilon) = n_0^{\rm MBB}\, \frac{8\,\pi}{(h c)^3}
        \frac{\varepsilon^2}{\mathrm{e}^{\frac{\varepsilon}{k T}}-1} \left(\frac{\varepsilon}{\varepsilon_0}\right)^\sigma
\label{app:eq:blackbody}
\end{equation}
where $T$ denotes the temperature in the case of pure black-body, and the absorption factor is given
by $(\frac{\varepsilon}{\varepsilon_0})^\sigma$ (see
e.g.~\cite{kruegel}).  $h$, $k$ and $c$ are the Planck constant,
Boltzmann constant and speed of light respectively.  For $\sigma=0$
and $n_0^{\rm MBB}=1$ the unmodified black-body spectrum is obtained.
The relation between the peak energy and temperature parameter is given
by a modified Wien's displacement law,
\begin{equation}
   \varepsilon_0 = \left[{\rm W}\!\left(- {\rm e}^{-b}\, b\right) + b\right] k \, T,
\end{equation}
where ${\rm W}(x)$ is the Lambert function (see e.g.~\cite{Veberic:2012ax}) and
$b = \sigma+2$.\\

For the study of the effect of using different functional forms of photon
spectra in our model (cf.\, Sec.\,\ref{sec:modelVarPhotonSpec}),
it is useful to use a common normalization for all spectra. The integral
photon density of Eq.\,(\ref{app:eq:photonfield}) is
\begin{equation}
  I_{\rm BPL} = n_0^{\rm BPL}\, \varepsilon_0 \left(\frac{1}{\alpha+1} - \frac{1}{\beta+1}\right)
\end{equation}
and for Eq.\,(\ref{app:eq:blackbody}) it is
\begin{equation}
  I_{\rm MBB}^\sigma = n_0^{\rm MBB}\,\frac{8\,\pi}{(h c)^3}\,(k T)^3\,\zeta(\sigma + 3, 1) \Gamma(\sigma + 3),
\end{equation}
where  $\zeta(x)$ denotes the Riemann zeta function and $\Gamma(x)$ is the Gamma function.
Choosing the  photon density of the unmodified black body spectrum as reference we use the
following normalization constants,
\begin{equation}
   n_0^{\rm BPL} = I_{\rm MBB}^0 / I_{\rm BPL}
\end{equation}
and
\begin{equation}
   n_0^{\rm MBB}(\sigma) = I_{\rm MBB}^0 / I_{\rm MBB}^\sigma.
\end{equation}
An example of the four photon spectra after normalization and for the same peak energy of
$\varepsilon_0 = 50$~\meV is shown in the left panel of Fig.\,\ref{app:photon-spectra}. The corresponding
interaction time for the sum of photo-dissociation and photo-pion production is shown in the right panel.

\section{Photo-Nuclear Interactions}
\label{app:appendixB}

The interaction between photons and high energy nuclei has been
extensively discussed in the
literature~\cite{Stecker:1969fw,Puget:1976nz,Karakula:1993he,Anchordoqui:1997rn,Epele:1998ia,Stecker:1998ib,Khan:2004nd,Hooper:2008pm,Aloisio:2008pp,Aloisio:2010he,Ahlers:2010ty}. In
this appendix, we describe how we implement the photon-nucleus
collisions in our analysis. The interaction time for a highly relativistic nucleus with energy $E
= \gamma A m_p$ (where $\gamma$ is the Lorentz factor) propagating
through an isotropic photon background with energy $\varepsilon$ and spectrum
$n(\varepsilon)$, normalized so that the total number density of photons is $\int n ( \varepsilon ) d \varepsilon$, is given by~\cite{Stecker:1969fw}
\begin{equation}
 \frac{1}{\tau_\mathrm{int}} = \frac{c}{2} \,\int_0^\infty
                 d\varepsilon \,\frac{n(\varepsilon)}{\gamma^2 \varepsilon^2}\, \int_0^{2\gamma\varepsilon}
                 d\varepsilon^\prime \, \varepsilon^\prime\, \sigma(\varepsilon^\prime),
\label{app:eq:interaction}
\end{equation}
where $\sigma(\varepsilon^\prime)$ is the photo-nuclear interaction
cross section of a nucleus of mass $A m_{p}$ by a photon of energy
$\varepsilon'$ in the rest frame of the nucleus.

Detailed tables of
$\sigma(\varepsilon^\prime)$ are available in
\CRP~\cite{Kampert:2012fi, Batista:2013gka}.  We use the numerical tools provided
at~\cite{crpropaTools} to calculate the interaction times for the
photon field given by Eqs.~(\ref{app:eq:photonfield}) and (\ref{app:eq:blackbody}).


For illustrative purposes, the cross section can be
approximated by a single pole in the narrow-width
approximation,
\begin{equation}
\sigma (\varepsilon') = \pi\,\,\sigma_{\rm res}\,\,  \frac{\Gamma_{\rm
  res}}{2} \,\,
\delta(\varepsilon' - \varepsilon'_{\rm res})\, ,
\label{sigma}
\end{equation}
where $\sigma_{\rm res}$ is the resonance peak, $\Gamma_{\rm res}$ its
width, and $\varepsilon'_{\rm res}$ the pole in the rest frame of the
nucleus.  The factor of $1/2$ is introduced to match the integral
(i.e. total cross section) of the Breit-Wigner and the delta
function~\cite{Anchordoqui:2006pe}.

The mean interaction time  is obtained  substituting Eq.\,(\ref{sigma}) into Eq.\,(\ref{app:eq:interaction}),
\begin{eqnarray}
  \frac{1}{\tau_{\rm int} (E)} & \approx & \frac{c\, \pi\,
    \sigma_{\rm res}
    \,\varepsilon'_{\rm res}\,
\Gamma_{\rm res}}{4\,
    \gamma^2}
  \int_0^\infty \frac{d \varepsilon}{\varepsilon^2}\,\,\, n(\varepsilon) \,\,\,
  \Theta (2 \gamma \varepsilon - \varepsilon'_{\rm res}) \nonumber \\
  & = & \frac{c \, \pi \, \sigma_{\rm res} \,\varepsilon'_{\rm res}\,
    \Gamma_{\rm res}}{4 \gamma^2}
  \int_{\epsilon'_{\rm res}/2 \gamma}^\infty \frac{d\varepsilon}{\varepsilon^2}\,\,
  n (\varepsilon)  \, .
 \label{A1}
\end{eqnarray}
Substituting (\ref{app:eq:photonfield}) into (\ref{A1}) yields:
\begin{equation}
\frac{1}{\tau_{\rm int} (E)} = \frac{1}{\tau_b}
\left\{\begin{array}{ll}  \,
(E_b / E)^{\beta +1} & ~ E \leq E_b  \\
(1-\beta)/(1-\alpha) \left[\left( E_b/E \right)^{\alpha +1} -
  \left(E_b/E\right)^2 \right] +
\left(E_b/E\right)^2 & ~ E > E_b
\end{array} \right. \, ,
\end{equation}
where
\begin{equation}
\tau_b = \frac{ 2 \ E_b \ (1-\beta)} {c \, \pi \
  \sigma_{\rm res} \, A \, m_p \ \Gamma_{\rm res}
   \ n_0} \quad {\rm and} \quad
E_b = \frac{\varepsilon'_{\rm res} \ A \ m_p}{2 \varepsilon_0} .
\end{equation}
The parameters characterizing the photo-disintegration cross section are:
$\sigma_{\rm res} \approx 1.45\times 10^{-27}~{\rm cm}^2 \, A$,
$\Gamma_{\rm res} = 8~{\rm MeV}$, and $\epsilon'_{\rm res} = 42.65
A^{-0.21} \, (0.925 A^{2.433})~{\rm MeV},$ for $A > 4$ ($A\leq
4$)~\cite{Karakula:1993he}. The parameters for the photo-pion
production cross section are: $\sigma_{\rm res} \simeq 5.0 \times
10^{-28}~{\rm cm}^2 \, A$, $\Gamma_{\rm res} = 150~{\rm MeV}$, and
$\varepsilon'_{\rm res} = (m_\Delta^2 - m_p^2)/(2 m_p) \simeq 340~{\rm
  MeV}$~\cite{Agashe:2014kda}.

\section{Propagation in the Source Environment}
\label{app:appendixC}

In our simple model we consider interactions
and escape of particles treating the source environment as a leaky box. If at a
given time, $t_0$, $N(t=t_0) = N_0 $ particles are injected at random into the
source environment, then the number of particles $N$ remaining in the
source at any later time $t$ changes as
\begin{equation}
 \frac{dN}{dt} = -\frac{1}{\tau_\mathrm{esc}} \,N -\frac{1}{\tau_\mathrm{int}}\,N,
 \label{app:eq:sourceprop1}
\end{equation}
where $\tau_\mathrm{esc}$ and $\tau_\mathrm{int}$ are the escape and interaction
times respectively. Integration yields the time evolution of $N$ as
\begin{equation}
   N(t) = N_0 \, \mathrm{e}^{-\frac{t-t_0}{\tau}},
\end{equation}
where
\begin{equation}
   \tau = \frac{\tau_\mathrm{esc}\tau_\mathrm{int}}{\tau_\mathrm{esc} + \tau_\mathrm{int}}.
\end{equation}
The total number of escaping particles is given by
\begin{equation}
   N_\mathrm{esc}  =  \int_{t_0}^\infty
   \frac{1}{\tau_\mathrm{esc}}\, N(t)\, dt
 =  N_0 \,\frac{\tau_\mathrm{int}}{\tau_\mathrm{esc} +
  \tau_\mathrm{int}}
 =  N_0 \,\frac{1}{1 + \tau_\mathrm{esc}/\tau_\mathrm{int}} \equiv N_0 \, f_\mathrm{esc}
\label{app:eq:nesc}
\end{equation}
and likewise the number of particles suffering interactions is
\begin{equation}
   N_\mathrm{int} = N_0 \,\frac{\tau_\mathrm{esc}}{\tau_\mathrm{esc} + \tau_\mathrm{int}}
                 = N_0 \,\frac{1}{1 + \tau_\mathrm{int} /
                   \tau_\mathrm{esc}} \equiv N_0 \, f_\mathrm{int} \,,
\label{app:eq:nint}
\end{equation}
with $N_\mathrm{esc} + N_\mathrm{int} = N_0$. As can be seen,
$N_\mathrm{esc}$ and $N_\mathrm{int}$ depend only on the ratio of the
escape and interaction times, but not on the absolute value of either
of them.

In the following we consider sources at steady state, i.e.\ sources
which are active long enough to justify integrating to infinity in
Eq.\,(\ref{app:eq:nesc}) and for which the injected flux equals the
escaping flux.  Interacting particles constitute the source for
secondary particles of lower mass number.

Since the particle trajectory in the source is treated as a random walk starting from a random position, the escape time of a
secondary does not depend on the time it was produced. Therefore we
can apply Eqs.~(\ref{app:eq:sourceprop1}) to~(\ref{app:eq:nint}) also to
the secondary particle production, which greatly simplifies the
equations with respect to previous analytic approaches that
had been developed for the extra-galactic propagation of cosmic-ray
nuclei~\cite{Hooper:2008pm, Ahlers:2010ty, Aloisio:2008pp, Aloisio:2010he}.

\subsection{Single-Nucleon Emission}

\label{app:sec:singlenuc}

The basic principle of the analytic calculation can be best
illustrated by firstly describing the case where interactions with the
photon field lead to the knock-out of a single nucleon,
\begin{equation}
   A + \gamma \rightarrow (A-1) + n/p,
\end{equation}
and the nucleon carries away a fraction of $1/A$ of the initial energy
of the nucleus. This approach has been successfully applied to the
photo-disintegration (PD) during the extra-galactic propagation of
nuclei (see e.g.~\cite{Hooper:2008pm}). It can also serve as a good
approximation for the losses due to photo-pion production (PP) if
nuclei are treated as the superposition of $A$ individual nucleons
(see e.g.~\cite{Kampert:2012fi, Batista:2013gka}).  The interaction
time is therefore the combination of the two processes, i.e.
\begin{equation}
   \tau_\mathrm{int} =
  \frac{\tau_\mathrm{int}^\mathrm{PD} + \tau_\mathrm{int}^\mathrm{PP}}
       {\tau_\mathrm{int}^\mathrm{PD} \tau_\mathrm{int}^\mathrm{PP}}.
\label{app:eq:tauint}
\end{equation}

In this simplified propagation scheme, secondaries with mass $A$ and
energy $E^*$ originate from nuclei with energy $E^\prime
= \frac{A+1}{A}\,E^*$ and mass $A+1$. They are produced at a rate
\begin{equation}
   \calQ(E^*,\, A) =
       \calQ_\mathrm{int}\left(\frac{A+1}{A}\, E^*, \, A+1\right)
       \,\left|\frac{dE^\prime}{dE^*}\right| =
       \calQ\left(\frac{A+1}{A}\, E^*, \, A+1\right)\,
       \eta_\mathrm{int}\left(\frac{A+1}{A}\, E^*,\, A+1\right) \,\frac{A+1}{A},
\label{app:eq:source}
\end{equation}
where the factor $\left|\frac{dE^\prime}{dE^*}\right|$ is
the Jacobian determinant needed to transform the differential
injection rate from the primary to secondary energy.  In analogy to
Eq.\,(\ref{app:eq:nesc}), a fraction of the
secondaries escapes the source environment,
\begin{equation}
 \calQ_\mathrm{esc}(E^*,\, A) =
    \calQ(E^*,\, A) \; \eta_\mathrm{esc}(E^*,\, A),
\end{equation}
and the remaining particles interact again at a rate of
\begin{equation}
 \calQ_\mathrm{int}(E^*,\, A) =
    \calQ(E^*,\, A) \; \eta_\mathrm{int}(E^*,\, A).
\end{equation}
This assumes that the escape probability of a secondary is independent
of the time or position it got produced in the source environment.
This calculation can be iterated to obtain the escape rate of any
remnant with mass $\Astar$ produced during the propagation of a nucleus of mass $A^\prime$:
\begin{equation}
  {\cal Q}_{\rm esc}^{\rm rem}(E^*,\, \Astar,A^\prime) =
     \calQ\left(\frac{A^\prime}{\Astar}\, E^*, \, A^\prime\right)\, \frac{A^\prime}{\Astar} \;
      \eta_\mathrm{esc}(E^*, \,\Astar) \,
     \prod_{A^\diamond = \Astar + 1}^{A^\prime} \eta_\mathrm{int}\left(\frac{{A^\diamond}}{\Astar}\, E^*, \, A^\diamond\right).
\label{app:QAesc}
\end{equation}
The rate of nucleons being knocked out of nuclei during propagation via
either of the considered processes $i = \rm PD / PP$ is
\begin{equation}
  {\cal Q} _{\rm esc}^i(E^*,\, n+p, A^\prime) =
     \calQ\left(\frac{A^\prime\, E^*}{\kappa_i}, \, A^\prime\right)\,
     \frac{A^\prime}{\kappa_i} \,\sum_{A^\pluscirc=2}^{A^\prime} \,
     f_i\left(\frac{A^\pluscirc\, E^*}{\kappa_i}, \, A^\pluscirc \right)
     \prod_{A^\diamond = A^\pluscirc}^{A^\prime} \eta_\mathrm{int}\left(\frac{A^\diamond\, E^*}{\kappa_i}, \, A^\diamond\right)
\label{app:Qpesc}
\end{equation}
with elasticities of the knock out nucleon given by $\kappa_{\rm PD} =
1$ and $\kappa_{\rm PP} = 0.8$ and the fractional contribution from PD
and PP given by
\begin{equation}
  f_{\rm PD} = \frac{1}{1 + \tau_\mathrm{PD}/\tau_\mathrm{PP}},
\quad {\rm and} \quad
 f_{\rm PP} = 1 - f_{\rm PD} = \frac{1}{1 + \tau_\mathrm{PP}/\tau_\mathrm{PD}}.
\end{equation}

The  total escape rate of particles of mass $\Astar$ from injected nuclei of mass  $A^\prime$ is
\begin{equation}
{\cal Q}_{\rm esc}^{\rm tot} (E^*,\Astar,A^\prime) =
{\calQ}_{\rm esc}^{\rm rem} (E^*,\Astar,A^\prime) +
\delta_{\!A^{\!^*}1} \left[ {\cal Q}_{\rm esc}^{\rm PD}
(E^*,\Astar,A^\prime) + {\cal Q}_{\rm esc}^{\rm PP}
(E^*,\Astar,A^\prime)\right] \,,
\label{app:eq:escTot}
\end{equation}
where $\delta_{A1}$ is the Kronecker delta.

\subsection{Branching Ratios from Photo-disintegration}
\label{app:sec:multinuc}
The propagation scheme described in the last section can be easily
extended to take into account the emission of several nucleons or
light nuclei in photo-nuclear reactions. We use the total interaction
time and branching ratios for photo-dissociation from
{\sc Talys}~\cite{talys, TALYSRestored} as available in \CRP and neglect multi-nucleon
emission for photo-pion production since it can be safely neglected
at the energies relevant here.

Instead of the closed formulae derived above for the single-nucleon
case, we now have the following recursive relation for the rate of
produced remnant nuclei of mass $\Astar$.
\begin{equation}
   {\cal Q}(E^*,\, \Astar) =
   \sum_{i=1}^{A^\prime - A^*} b(E_i,\, \Astar, A_i) \,\eta_\mathrm{int}\left(E_i,\, A_i\right)
    {\cal Q}(E_i,\, A_i) \,\left|\frac{d E_i}{dE^*}\right|
\label{app:eq:multinuc}
\end{equation}
with $A_i = \Astar + i$, $E_i= A_i/\Astar\, E^*$ and $\left|\frac{d
  E_i}{dE^*}\right| = A_i/\Astar$. $b(E_i,\, \Astar, A_i)$ is the
branching ratio that gives the probability that a nucleus of mass
$A_i$ with energy $E_i$ will have a remnant mass of $\Astar$ after the
interaction. The $n$ knocked out nucleons and nuclei are calculated
the same way but replacing the branching fraction by $n\, b(E_i,\, n\,
\Astar, A_i)$, i.e. the probability to produce $n$ fragments of mass
$\Astar$, and summing over $n$. The rest of the calculation proceeds
as in the case of single-nucleon emission.

\subsection{Proton Interactions}
\label{app:sec:proton}
Once nucleons are generated in photo-disintegration they are assumed
to either escape immediately in the case of neutrons, or to interact
further via photo-pion production. The average elasticity of this
process is $\kappa_{\rm PP} = 0.8$ and corresponds to a shift in
energy of $\Delta\lg E = \lg \kappa_{\rm PP} \approx 0.1$. Since we
perform the calculation in logarithmic bins of this width, proton
interactions can be treated similarly to photo-disintegration as a
``trickle-down'' of particles fluxes subsequently shifted by one
energy bin. For the reaction $p+\gamma \rightarrow \pi^+ + n$, the
neutron escapes and the interaction chain is finished. In case of
$p+\gamma \rightarrow \pi^0 + p$, the secondary proton has a reduced energy; it may also
interact again. The neutron, proton and positive pion fluxes in
an energy bin $k$ are calculated from the recursive relations
\begin{equation}
   {\mathcal Q}(E^*_k, n)^\prime = {\mathcal Q}(E^*_k,\, n) + (1-b_{pp}) \; \eta_{\rm int}(E^*_{k+1},\, p)
                                  \,  {\mathcal Q}(E^*_{k+1}, p)^\prime / \kappa
\end{equation}
\begin{equation}
   {\mathcal Q}(E^*_k, p)^\prime = {\mathcal Q}(E^*_k,\, p) + b_{pp} \; \eta_{\rm int}(E^*_{k+1},\, p)
                                  \,  {\mathcal Q}(E^*_{k+1}, p)^\prime / \kappa
\end{equation}
and
\begin{equation}
   {\mathcal Q}(E^*_k, \pi^+)^\prime = (1-b_{pp}) \;\eta_{\rm int}(E^*_{k+7},\, p)
                                  \,  {\mathcal Q}(E^*_{k+7}, p)^\prime / (1-\kappa),
\end{equation}
where $b_{pp} \approx 0.5$ is the branching fraction of the process $p+\gamma
\rightarrow \pi^0 + p$ and the un-primed fluxes are the sum of the
knocked-out nucleons from Eq.\,(\ref{app:eq:multinuc}) and primary
protons. The offset of 7 in the equation for the pion flux is due
to the energy shift of the pions, $\Delta\lg E = \lg (1-\kappa_{\rm PP}) \approx 0.7$.

\section{Cosmic Ray Production and Propagation in an Expanding Universe}
\label{app:appendixD}

To compare the spectra obtained in the last section, the particles
need to be propagated to Earth.  
The number of cosmic rays per unit volume and energy in the present universe is equal to the
number of particles accumulated during the entire
history of the universe and is comprised of both primary
particles emitted by the sources and secondaries produced in the
photo-disintegration process.  Herein, the variable $t$ characterizes
a particular age of the universe and $t_H$ indicates its present
age. We adopt the usual concordance cosmology of a flat universe
dominated by a cosmological constant, with $\Omega_\Lambda \approx
0.69$ and a cold dark matter plus baryon component $\Omega_m \approx
0.31$~\cite{Ade:2015xua}. The Hubble parameter as a function of
redshift $z$ is given by $H^2(z) = H_0^2 [\Omega_m (1 + z)^3 +
  \Omega_\Lambda]$, normalized to its value today, $H_0 = 100 \,
h~{\rm km} \ {\rm s}^{-1} \, {\rm Mpc}^{-1}$, with $h \simeq
0.68$~\cite{Ade:2015xua}. The dependence of the cosmological time with
redshift can be expressed via $dz = - dt (1 + z) H(z)$. The co-moving
space density of cosmic rays $n_{\rm CR}$ of mass $A$ from a population of
uniformly distributed sources with (possibly age-dependent) emission rate per
volume $\calQ(E^\prime,A^\prime,t)$ is given by
\begin{eqnarray}
n_{\rm CR}  (E,A,A^\prime)  \equiv
\frac{dN_{\rm CR}}{dE\,dV}
 =   \int_E^\infty \! \!  \int_0^{t_H}
\!\! \frac{d\mathscr{P}_{AA^\prime}  (E^\prime, E, t)}{dE} \
\calQ(E^\prime,A^\prime,t) \  \xi(t)  \ dE^\prime \ dt \,,
\label{app:nCR}
\end{eqnarray}
where $d\mathscr{P}_{AA^\prime}/dE$ is the expectation value for the
 number of nuclei of mass $A$ in the energy interval ($E, E+dE)$ which
 derive from a parent of mass $A^\prime$ and energy $E^\prime$ emitted
 at time $t$, and $\xi(t)$ is the ratio of the product of co-moving
 source density and $\calQ(E^\prime,A^\prime,t)$, relative to the
 value of that product today. Note that $d\mathscr{P}_{AA^\prime}/dE$
 includes propagation effects both at the source environment and {\it
 en route} to Earth.

We assume that the emission rate of cosmic rays is the same for all
sources and the spectrum and composition is independent of the age of the universe, so that evolution of the volumetric emission rate with cosmological time can be described by an overall source evolution factor, $\xi(t)$ discussed below.  (It need not be specified whether this is due to an evolution of the number of sources or their intrinsic power.)  We further
assume, as per usual practice, that emission rate is fairly well
described by a power-law spectrum. Under these general assumptions the
source emission rate per volume takes the form
\begin{equation}
\calQ(E^\prime,A^\prime) = \calQ_0 \left(\frac{E^\prime}{E_0}\right)^{\gamma}
\exp\left(-\frac{E^\prime}{Z' E^p_{\rm max}}\right),
\label{app:injectionQ}
\end{equation}
where $E^p_{\rm max}$
is the maximal energy of emitted protons, i.e., maximum rigidity of the accelerator, $Z'$ is the nucleus' atomic
number, $E_0$ is some reference energy, and
\begin{equation}
 \calQ_0 = \left\{\begin{array}{rl}  \bolddot{n}_0  \left.  \frac{dN_{A^\prime}}{dE^\prime}\right|_{E^\prime= E_0},  & ~~{\rm for \
       bursting \ sources} \\
n_0  \left. \frac{dN_{A^\prime}}{dE^\prime dt}\right|_{E^\prime=E_0},  &
~~{\rm for \ steady \ sources}
\end{array} \right. .
\end{equation}
Here, $\bolddot{n}_0$ is the number of bursts per unit volume per unit
time and $dN_{A^\prime}/dE^\prime$ is the spectrum of particles produced by each burst, or for a steady source $n_0$ is the number density of sources at $z=0$, and $dN_{A^\prime}/dE^\prime dt$ is the UHECR production rate per unit energy per source.  The cosmic ray power
density above a certain energy $E^\prime_{\rm min}$ is given by
\begin{eqnarray}
    &&\bolddot{\epsilon}_{E^\prime}(A^\prime) = \int_{E^\prime_{\rm
        min}}^\infty E^\prime \, \calQ(E^\prime,A^\prime) \, dE^\prime \nonumber \\
   & =& Q_0 \int_{E^\prime_{\rm min}}^\infty
        E^\prime \left(\frac{E^\prime}{E_0}\right)^{\gamma}
        \exp\left(-\frac{E^\prime}{Z^\prime E^p_{\rm max}}\right)\,
        dE^\prime \nonumber \\
   & =& Z^\prime E^p_{\rm max} \left(\frac{Z^\prime E^p_{\rm max}}{E_0}\right)^{\gamma+1}
       \int_{E^\prime_{\rm min} / (Z^\prime E^p_{\rm max})}^\infty
      t^{\gamma+1} {\rm e}^{-t}\, dt \nonumber \\
   & =& \calQ_0\, E_0^2 \left(\frac{Z^\prime E^p_{\rm max}}{E_0}\right)^{\gamma+2}
        \Gamma\left(\gamma+2,\, \frac{E^\prime_{\rm min}}{Z^\prime
            E^p_{\rm max}}\right),
\end{eqnarray}
where $\Gamma$ denotes the upper incomplete gamma function.

The cosmological evolution of the source density per co-moving volume
is parametrized as
\begin{equation}
 n_{\rm s} (z) = n_0\, \xi(z)
\end{equation}
with $\xi(z=0)=1$.
We adopt for the fiducial model that the evolution of sources
follows the star formation rate with
\begin{equation}
\xi(z) =  \frac{(1+z)^a}{1 + [(1+z)/b]^{c}}
\label{eq:evolution}
\end{equation}
where $a = 3.26 \pm 0.21$, $b = 2.59 \pm 0.14$ and $c = 5.68 \pm
0.19$~\cite{Robertson:2015uda}.  Additionally we consider the family of evolution models parameterized as $\xi(z) =  (1+z)^{m}$.

To propagate the particles escaping the source environment to Earth we
use the \CRP framework~\cite{Kampert:2012fi, Batista:2013gka}.  For
this purpose, we generate a library of propagated nuclei with $\Astar
= 1 \dots \Astar_\mathrm{max}$ injected uniformly in light-travel
distance. The latter corresponds to a non-evolving source distribution
in comoving distance after accounting for the cosmological time
dilation. We simulated particles up to $\Astar_\mathrm{max}=56$.
Given this library of simulated particles, we can construct the
propagation matrix ${\cal M}_{i j \mu \nu}$ for arbitrary source
evolutions for each nuclear mass $A^*_\mu$ escaping the source and
secondary mass $A_\nu$ at Earth. The elements of the propagation
matrix give the expected number of secondaries in an energy interval
$[\lg E_j, \lg E_j+\Delta]$ at Earth originating from nuclei at the
source at an energy $[\lg E_i^*, \lg E_i^*+\Delta]$ for a given source
evolution $\xi(z(t))$ and a uniform logarithmic spacing in energy with
$\Delta = 0.1$. Numerically, the elements are constructed via
discretization of Eq.\,(\ref{app:nCR})
\begin{equation}
 n_{\rm CR} (E_j,A_\nu,A')  =  \sum_{A^*_\mu=A_\nu}^{A'} \ \sum_{i=j}^{n_i} \
\sum_{a=0}^{n_a} \frac{\Delta {\cal P}_{i j \mu \nu a}}{\Delta E_j} \
  {\cal Q}_{\rm esc}^{\rm tot} (E^*_i,A^*_\mu,A') \ \xi(t_a)  \ \Delta t_a \ \Delta E^*_i \,,
\label{app:Auno}
\end{equation}
where $\Delta t_a = t_H/n_a$ and
\begin{equation}
\frac{\Delta {\cal P}_{i j \mu \nu a}}{\Delta E_j} = \frac{1}{\Delta E_j}
\frac{N^{\rm Earth}_{i j \mu \nu a}( E^*_i, E^*_i + \Delta E^*_i;
  E_j, E_j + \Delta E_j;
A^*_\mu;A_\nu;t_a, t_a + \Delta t_a)}{N^{\rm gen}_{i \mu a} (E^*_i, E^*_i +
  \Delta E^*_i;A^*_\mu; t_a, t_a+\Delta t_a)}
\label{app:Ados}
\end{equation}
For a non-evolving injection rate per unit volume, the number of
generated events per bin is constant, $N^{\rm gen}_{i \mu a} = K^{\rm
gen}_{i \mu}$.  Then, for any source evolution $\xi[z(t))]$,
(\ref{app:Auno}) can be rewritten as
\begin{eqnarray}
n_{\rm CR} (E_j,A_\nu,A')  & = &\sum_{A^*_\mu = A_\nu}^{A'} \sum_{i =j}^{n_i}
{\cal Q}_{\rm esc}^{\rm tot} (E^*_i, A_\mu^*, A') \ \frac{\Delta
  E^*_i}{\Delta E_j}  \ \sum_{a=0}^{n_a}
\frac{N_{i j \mu \nu a}^{\rm Earth}}{N^{\rm gen}_{i \mu a} } \ \xi[z(t_a)] \
 \ \Delta t_a \nonumber
\\
 & = & \sum_{A^*_\mu = A_\nu}^{A'} \sum_{i =j}^{n_i}
{\cal Q}_{\rm esc}^{\rm tot} (E^*_i,A^*_\mu, A_\nu) \frac{\Delta E^*_i}{\Delta E_j}
\ t_H \ \frac{ \sum_{a=0}^{n_a}  N_{i j \mu \nu a}^{\rm Earth} \ \xi[z(t_a)]}{n_a \ K^{\rm
    gen}_{i \mu}}  \nonumber \\
 & = &
\sum_{A^*_\mu = A_\nu}^{A'} \sum_{i =j}^{n_i}
{\cal Q}_{\rm esc}^{\rm tot} (E^*_i,A^*_\mu,A') \ \frac{\Delta
  E^*_i}{\Delta E_j} \ t_H \
\frac{ \sum_{a=0}^{n_a}  N_{i j \mu \nu a}^{\rm Earth} \ \xi [z(t_a)]}{\sum_{a=0}^{n_a}  N^{\rm
    gen}_{i \mu a}}  \nonumber \\
& = & \sum_{A^*_\mu = A_\nu}^{A'} \sum_{i =j}^{n_i}
{\cal Q}_{\rm esc}^{\rm tot} (E^*_i,A^*_\mu,A') \ \frac{\Delta
  E^*_i}{\Delta E_j} \ t_H \
\frac{\sum_{p=0}^{N^{\rm Earth}_{i j \mu \nu}} \xi[z(t_p)] }{ N^{\rm
    gen}_{i \mu}} \nonumber \\
& = & \sum_{A^*_\mu = A_\nu}^{A'} \sum_{i =j}^{n_i}
\frac{\Delta
  E^*_i}{\Delta E_j} \ t_H \ {\cal M}_{i j \mu \nu} \,  {\cal
  Q}_{i \mu} \,,
\label{app:Atres}
\end{eqnarray}
where $\sum_{p=0}^{N^{\rm Earth}_{i j \mu \nu}} \xi[z(t_p)]$ denotes
the $\xi$-weighted sum over all events generated with $(A^*_\mu,
E^*_i)$ arriving at Earth with $(A_\nu, E_j)$ and   $N^{\rm
    gen}_{i \mu}$ is the total number of generated events with
  $(A^*_\mu, E_i)$.  Note that if binned in $\Delta z_b =
    z_{\rm max}/b$, then $N^{\rm gen}_{i \mu b}  |\Delta
z_b/\Delta t_b| =\,{\rm constant}$, and hence (\ref{app:Atres}) can be rewritten as
\begin{equation}
n_{\rm CR} (E_i,A_\nu,A')  =
\sum_{A^*_\mu = A_\nu}^{A'} \sum_{i =j}^{n_i}
{\cal Q}_{\rm esc}^{\rm tot} (E^*_i,A^*_\mu,A')  \frac{\Delta
  E^*_i}{\Delta E_j}  z_{\rm max}
\frac{ \sum_{a=0}^{n_a}  N_{i j \mu \nu a}^{\rm Earth}  \xi(z_b)}{\sum_{a=0}^{n_a}  N^{\rm
    gen}_{i \mu a}  \left|\frac{\Delta z_a}{\Delta t_a}\right|},
\end{equation}
where $|\Delta
z_b/\Delta t_b| = (1 + z_b) H(z_b)$ and $z_{\rm max} = \Delta z_b \,
n_b$.

For a given spectrum of injected nuclei of mass $A^\prime$
we obtain the space density of cosmic rays at Earth with
energy $E$ and mass $A$,
\begin{equation}
n_{\rm CR}  (E,A,A^\prime)  =
\frac{dN_{\rm CR}}{dE\,dV}.
\end{equation}
For an isotropic arrival direction distribution (which is an excellent
approximation based on current observations) the relation between the
spectrum and the cosmic ray density  is
\begin{equation}
J (E, A, A^\prime) \equiv  \frac{dN_{\rm CR}}{dE \ dA \ dt \ d\Omega}
  =  \frac{c}{4\pi} n_{\rm CR} (E,A,A^\prime).
\end{equation}
The total flux at earth of particles of mass $A_\nu$ is
\begin{equation}
J (E, A_\nu) = \sum_{A_\mu^\prime = A_\nu}^{A^\prime_{\rm max}} f(A_\mu^\prime) J (E, A_\nu, A_\mu^\prime),
\end{equation}
where $f(A_\mu^\prime)$ denotes the fraction of particles of mass
$A_\mu^\prime$ injected at the source.

\section{Neutrino and Photon Production}
\label{app:appendixE}

The results of the last two sections can be readily applied to obtain
the flux of neutrinos at Earth from the decay of neutrons and charged
pions. We approximate the emission rate of pions from photo-pion
production by using $\kappa_\pi = 1 - \kappa_{\rm PP}$ in Eq.\,(\ref{app:Qpesc}).  The
energies of neutrinos escaping from the source are given by the
kinematics of the two-body decay of pions and the subsequent muon
decay which we treat approximately by assigning a third of the muon
energy to each of the decay products. In this way we construct a
propagation matrix for pions, ${\cal M}_{i j \mu \nu}$ with $_\nu =
(\nu_\mu, \bar{\nu}_\mu, \nu_e (\bar{\nu}_e))$ and $_\mu=\pi^\pm$.
Similarly, a propagation matrix for neutrons is obtained with $_\nu =
(\bar{\nu}_e)$ and $_\mu=n$.

Neutrino oscillation over astronomical distances modifies the initial
flavor distribution of fluxes, $\Phi^0_e : \Phi^0_\mu : \Phi^0_\tau$,
in calculable ways. The  relevant  parameters for such a
calculation are the three Euler rotations ($\theta_{12}$,
$\theta_{23}$, $\theta_{13}$) and the $CP$-violating Dirac phase $\delta$.  The current best-fit
values as well as the allowed ranges of the mixing parameters at
the $1\sigma$ level are: $\theta_{12}/^\circ =
33.57^{+0.77}_{-0.75}$, \mbox{$\theta_{23}/^\circ = 41.9^{+0.5}_{-0.4}
\oplus 50.3^{+1.6}_{-2.5}$},  $\theta_{13}/^\circ =
8.73^{+0.35}_{-0.36}$, $\delta/^\circ = 266^{+55}_{-55}$~\cite{Gonzalez-Garcia:2014yaa}.
The mixing probabilities are given by
\begin{equation}
P_{\nu_\mu \to \nu_\mu} =  c_{13}^4 s_{23}^4 + (c_{12}^2 c_{23}^2 +
s_{12}^2 s_{13}^2 s_{23}^2 -
  2 c_{12} c_{23} s_{12} s_{13} s_{23} c_\delta)^2 +
(c_{23}^2 s_{12}^2 + c_{12}^2 s_{13}^2 s_{23}^2 + 2 c_{12} c_{23}
s_{12} s_{13} s_{23} c_\delta)^2 \,,
\end{equation}
\begin{equation}
P_{\nu_e \leftrightarrow \nu_\mu}
=   2 \, c_{13}^2 \,
\left\{c_{12}^2 \, s_{12}^2 \, c_{23}^2
+ \left(c_{12}^4+ s_{12}^2 \right) \, s_{13}^2 \, s_{23}^2
 + c_{12} \, s_{12} \,
    c_{23} \, s_{23} \, c_\delta \, (c_{12}^2 - s_{12}^2)
    \, s_{13}
 \right\} ,
\end{equation}
\begin{equation}
 P_{\nu_e \leftrightarrow \nu_\tau} = P_{\nu_e \leftrightarrow
   \nu_\mu} (\theta_{23} \to \theta_{23} + \pi/2) \,,
\quad
P_{\nu_\tau \to \nu_\tau} = P_{\nu_\mu \to
   \nu_\mu} (\theta_{23} \to \theta_{23} + \pi/2) \,,
\end{equation}
and the unitarity relations
\begin{equation}
P_{\nu_e \to \nu_e} = 1 - P_{\nu_e \leftrightarrow \nu_\mu} - P_{\nu_e
  \leftrightarrow \nu_\tau} \,
\quad
P_{\nu_\tau \leftrightarrow \nu_\mu} = 1 - P_{\nu_e \leftrightarrow \nu_\mu} - P_{\nu_\mu
  \to \nu_\mu} \,
\quad
P_{\nu_\tau \to \nu_\tau} = 1 - P_{\nu_e \leftrightarrow \nu_\tau} - P_{\nu_\mu
  \leftrightarrow \nu_\tau} \ ,
\end{equation}
where $c_{ij} = \cos \theta_{ij}$, $s_{ij} = \sin \theta_{ij}$, and
$c_\delta = \cos \delta$~\cite{Pakvasa:2007dc}.
The measurable neutrino flux at Earth is given by
\begin{equation}
\left(\begin{array}{c}\Phi_e \\ \Phi_\mu \\ \Phi_\tau \end{array} \right) =
 \left(\begin{array}{ccc} 0.55 & 0.24 & 0.21 \\ 0.24 &
    0.37 & 0.38 \\ 0.21 & 0.38 & 0.41 \end{array}\right) \,
\left( \begin{array}{c} \Phi^0_e  \\ \Phi^0_\mu \\
    \Phi^0_\tau \end{array} \right) \, .
\end{equation}

In addition to neutrinos, photons are produced from
$\pi^0$ production and decay~\cite{Aharonian:2004yt}, and by photo-disintegration of
high-energy nuclei followed by immediate photo-emission from the
excited daughter nuclei~\cite{Anchordoqui:2006pd}.
The $\gamma$-rays, electrons, and positrons produced in the decay of
$\pi^0$ and $\pi^\pm$ trigger an electromagnetic (EM) cascade on the cosmic microwave
background, which develops via repeated $e^+e^-$ pair production and
inverse Compton scattering. Other contributions to the cascade are
provided by Bethe-Heitler production of $e^+e^-$ pairs and
$\gamma$-rays emitted during the photo-disintegration process, after
the photo-dissociated nuclear fragments de-excite. The net result is a
pile up of $\gamma$-rays at ${\rm GeV} \alt E_\gamma \alt {\rm TeV}$,
just below the threshold for further pair production on the diffuse
optical backgrounds. The EM energy then gets recycled into the
so-called Fermi-LAT region, which is bounded by
observation~\cite{Abdo:2010nz,Ackermann:2014usa} to not exceed
$\omega_{\rm cas} \sim 5.8 \times 10^{-7}~{\rm
 eV/cm^3}$~\cite{Berezinsky:2010xa}.   The latest Fermi-LAT limts \cite{Ackermann:2014usa} do not significantly influence the determination of the $\omega_{\rm cas}$ upper bound of ~\cite{Berezinsky:2010xa}, because that bound is not very sensitive to the high energy bins added by ~\cite{Ackermann:2014usa}.

The photons coming from photo-pion production in the source environment were shown to be below the Fermi-LAT bound in Section \ref{sec:fidmodel}.  To place a bound on the contribution of photons from nuclear de-excitation to the Fermi-LAT diffuse gammas, without performing an explicit calculation, we can turn to the estimate of ~\cite{Anchordoqui:2014pca} which found $\omega_{\rm cas} \approx 1.1 \times 10^{-7}~{\rm
  eV/cm^3}$ assuming the high-energy IceCube spectrum to be entirely due to neutron beta-decay.  Since in our model the neutrino flux from neutron decay is significantly below the IceCube spectrum, the corresponding de-excitation photon contribution to the Fermi-LAT data must be far below the limit.

\twocolumngrid

%

\end{document}